\documentclass[twocolumn,aps,prd,showpacs,nofootinbib,superscriptaddress,preprintnumbers,floatfix,10pt]{revtex4-1}

\usepackage{amsmath}
\usepackage{amsfonts}
\usepackage{amssymb}
\usepackage{epsfig}
\usepackage{graphicx}
\usepackage{color}
\usepackage{slashed}
\usepackage{epstopdf}
\usepackage{dsfont}

\usepackage{hyperref}

\newcommand{\tr}{\text{tr}}

\newcommand{\Tr}{\text{Tr}}
\newcommand{\Eqref}[1]{Eq.~\eqref{#1}}

\newcommand{\Nf}{N_{\mathrm{f}}}

\newcommand{\yb}{\bar{\psi}}
\newcommand{\dir}{\slashed{\nabla}}
\newcommand{\kir}{k_{\text{IR}}}

\newcommand{\Aa}{\mathcal{A}}
\newcommand{\Nfgc}{N_{\text{f,gc}}}
\newcommand{\blamc}{\bar{\lambda}_{\text{cr}}}
\newcommand{\blamL}{\bar{\lambda}_\Lambda}

\begin{document}

\preprint{}

\title{A curvature bound from gravitational catalysis} 

\author{Holger Gies}
\email{holger.gies@uni-jena.de}
\affiliation{Theoretisch-Physikalisches Institut, 
Abbe Center of Photonics, Friedrich Schiller University Jena, Max Wien 
Platz 1, 07743 Jena, Germany}
\affiliation{Helmholtz-Institut Jena, Fr\"obelstieg 3, D-07743 Jena, Germany}
\author{Riccardo Martini}
\email{riccardo.martini@uni-jena.de}
\affiliation{Theoretisch-Physikalisches Institut, 
Abbe Center of Photonics, Friedrich Schiller University Jena, Max Wien 
Platz 1, 07743 Jena, Germany}

\begin{abstract}
  We determine bounds on the curvature of local patches of spacetime
  from the requirement of intact long-range chiral symmetry. The
  bounds arise from a scale-dependent analysis of gravitational
  catalysis and its influence on the effective potential for the
  chiral order parameter, as induced by fermionic fluctuations on a
  curved spacetime with local hyperbolic properties. The bound is
  expressed in terms of the local curvature scalar measured in units
  of a gauge-invariant coarse-graining scale. We argue that any
  effective field theory of quantum gravity obeying this curvature
  bound is safe from chiral symmetry breaking through gravitational
  catalysis and thus compatible with the simultaneous existence of
  chiral fermions in the low-energy spectrum.  With increasing number
  of dimensions, the curvature bound in terms of the hyperbolic scale
  parameter becomes stronger. Applying the curvature bound to the
  asymptotic safety scenario for quantum gravity in four spacetime
  dimensions translates into bounds on the matter content of particle
  physics models.
\end{abstract}


\maketitle

\section{Introduction}
\label{intro}

Gravitational catalysis denotes the breaking of chiral symmetry and
subsequent fermionic mass generation induced by a curved spacetime
background. The phenomenon is known to occur generically in fermionic
systems of any dimension for various negatively curved spacetimes even
at the weakest fermionic attraction \cite{Buchbinder:1989fz,Buchbinder:1989ah,Buchbinder:1989ah,Inagaki:1993ya,Sachs:1993ss,Elizalde:1995kg,Kanemura:1995sx,Inagaki:1995bk,Inagaki:1996nb,Geyer:1996wg,Geyer:1996kg,Miele:1996rp,Vitale:1998wm,Inagaki:1997kz,Hashida:1999wb,Gorbar:2007kd,Hayashi:2008bm,Inagaki:2010py,Sasagawa:2012mn,Gorbar:1999wa,Ebert:2008pc}.

Gravitational catalysis can be understood as a consequence of
dimensional reduction of the fluctuation spectrum. For instance in
$D$-dimensional hyperbolic space, the low-lying modes of the Dirac
operator exhibit a reduction from $D$ to $1+1$ dimensions
\cite{Gorbar:2008sp}. Hence, the long-range dynamics of any
self-interaction of the fermions (be it fundamental, effective or
induced) involving a chiral symmetry-breaking channel behaves like the
corresponding model in $1+1$ dimensions, e.g., the Gross-Neveu or the
Nambu--Jona-Lasinio model, which both exhibit chiral symmetry
breaking and fermionic mass generation.

In this respect, gravitational catalysis is closely related to
magnetic catalysis
\cite{Gusynin:1994va,Gusynin:1994xp,Gusynin:1995nb,Klevansky:1989vi,Klimenko:1990rh,Shovkovy:2012zn}
of chiral symmetry breaking in a magnetic field where a dimensional
reduction mechanism is also visible in the fermionic fluctuation
spectrum in the form of the lowest Landau level. Both phenomena can
also be understood within a renormalization group framework
\cite{Scherer:2012nn,Gies:2013dca}, where an analysis of the RG flow
reveals that the chiral channels inevitably become relevant operators
even in higher dimensions, once the long-range flow is driven by the
low-lying modes of the fermion spectrum.

Unlike the case of magnetic catalysis, the relevance of gravitational
catalysis for real systems is less clear. While the mechanism still
works in negatively curved space (instead of spacetime) such as on the
Lobachevsky plane \cite{Gorbar:2007kd,Gies:2013dca}, an estimate for
the required curvature for inducing a chiral transition in layered
materials with Dirac fermionic excitations, i.e., a Mott transition,
results in large negative values which seem difficult to achieve with
current materials \cite{Gies:2013dca}.

In the present work, we argue that gravitational catalysis may play a
malign role for the interplay of quantum gravitational and fermionic
matter degrees of freedom in the high-energy regime near the Planck
scale. As suggested in \cite{Eichhorn:2011pc}, the observational fact of the
existence of light chiral fermions in our universe puts implicit
bounds on the properties of the quantum gravitational interactions: if
quantum gravity near the Planck scale was such that it triggered
chiral symmetry breaking, the low-energy particle sector of our
universe would generically be characterized by massive fermions with
Planck scale masses. As gravity couples equally to all matter degrees
of freedom, it thus would seem difficult to understand the existence
of light chiral fermions.

Despite the fact that gravity represents an attractive interaction
among particles, gravitational fluctuations in a quantum-field theory
setting surprisingly do not trigger chiral symmetry breaking
\cite{Eichhorn:2011pc,Meibohm:2016mkp}. In this respect, gravity
differs substantially from gauge theories or Yukawa
interactions. Therefore, the existence of light fermions appears
compatible even with a high-energy regime of strongly coupled gravity,
as long as an effective field theory description for quantum gravity
is suitable. In particular, the asymptotic safety scenario of quantum
gravity
\cite{Weinberg:1976xy,Weinberg:1980gg,Reuter:1996cp,Reuter:2001ag,Niedermaier:2006wt,Reuter:2012id,Percacci:2017book}
passes this consistency test
\cite{Dou:1997fg,Percacci:2002ie,Zanusso:2009bs,Vacca:2010mj,Dona:2013qba,Meibohm:2015twa,Oda:2015sma,Eichhorn:2016esv,Hamada:2017rvn,Biemans:2017zca,Eichhorn:2017eht},
also in conjunction with further gauge interactions
\cite{Daum:2009dn,Folkerts:2011jz,Christiansen:2017gtg,Christiansen:2017cxa}. In
asymptotic safety, it is even possible to study the interplay of
non-chiral fermions and gravity \cite{Eichhorn:2016vvy}, demonstrating
that chiral fermions are favored by simple asymptotic safety
scenarios. Certain asymptotically safe gravity-matter scenarios even
exhibit an enhanced predictive power
\cite{Shaposhnikov:2009pv,Harst:2011zx,Bezrukov:2012sa,Eichhorn:2017ylw,Eichhorn:2017lry,Eichhorn:2017egq,Eichhorn:2017muy,Wetterich:2016uxm}.

Whereas most of these studies have essentially been performed on flat
space, with curvature dependent calculations coming up only recently
\cite{Knorr:2017fus,Christiansen:2017bsy}, the gravitational catalysis
mechanism is active on negatively curved spacetimes. In this picture,
the consistency of quantum gravity and light fermions thus is not so
much a matter of gravitational fluctuations and their interplay with
matter, but of the effective spacetime resulting from quantum gravity
itself.

In order to elucidate the mechanism by which gravitational catalysis
can affect the realization of quantum gravity and its interplay with
the particle content of Nature, we perform a scale-dependent analysis
of gravitational catalysis. We introduce an infrared (IR) scale $\kir$
that serves as a coarse-graining scale for the fermionic long-range
modes that drive chiral symmetry breaking. Simultaneously, this scale
can be viewed as an inverse length scale of a local patch of spacetime
characterized by an averaged curvature. The relevance or irrelevance
of gravitational catalysis then arises as a competition between the
local curvature-induced contributions and the screening of
contributions from the long-range modes. This results in bounds on the
local curvature measured in units of the coarse-graining scale: in
order to evade fermion mass generation and chiral symmetry breaking,
the curvature bound has to be satisfied on all scales $\kir$.

This statement may become particularly relevant for a high-energy
scale of quantum gravity, where $\kir$ may be of the order of the
Planck scale. If a quantum gravity scenario violates our curvature
bound in the Planck regime, the possible onset of gravitational
catalysis can give rise to a fermion mass spectrum of the
corresponding particle physics sector which is expected to be of
Planck scale as well. Hence, a violation of the bound can be
indicative for a tension between a quantum gravity scenario and the
existence and observation of light chiral fermions in our universe.

Our paper is organized as follows: Sect.~\ref{sec:framework} lays out
the general framework of our study in terms of a generic chiral
fermion theory in curved spacetime, which we analyze in a local
mean-field RG approach. We illustrate our approach with the fully
analytically accessible simplest case of $D=3$ dimensional
spacetime. The most relevant $D=4$ dimensional case is analyzed in
Sect.~\ref{sec:fourDim}. Higher-dimensional cases are studied in
Sect.~\ref{sec:higherDim}, where we find that the curvature bound gets
stronger with increasing dimensionality. We illustrate the usefulness
of the curvature bound with the aid of the asymptotic safety scenario
for quantum gravity in a simple setting in Sect.~\ref{sec:applic}. In
this scenario, the curvature bound can, for instance, translate into a
bound on the admissible number of fermion flavors. We conclude in
Sect.~\ref{sec:conc}.

\section{Framework}
\label{sec:framework}

Let us start from a fermionic matter sector with a global chiral
symmetry group $U(\Nf)_{\text{R}}\times U(\Nf)_{\text{L}}$, with $\Nf$
being the number of fermion species. This is reminiscent to the
fermionic sector of the standard model subject to the strong
interaction with $\Nf$ counting the number of flavors times the number
of colors. Even without any further gauge interactions, gravitational
fluctuations, say in the (trans-)Planckian regime, will induce
effective fermionic self-interactions. With gravity preserving chiral
symmetry, the most general local fermionic self-interaction to
fourth-order in the fields is parametrized by the action \cite{Gies:2003dp,Eichhorn:2011pc}
\begin{align}
S[\yb,\psi]=& \int_x\;\Big\{\yb\dir\psi\notag\\
&+\frac{\bar{\lambda}_-}{2}\Big[\Big(\yb^a\gamma_{\mu}\psi^a\Big)^2+\Big(\yb^a\gamma_{\mu}\gamma_5\psi^a\Big)^2\Big]\notag\\
    &+\frac{\bar{\lambda}_+}{2}\Big[\Big(\yb^a\gamma_{\mu}\psi^a\Big)^2-\Big(\yb^a\gamma_{\mu}\gamma_5\psi^a\Big)^2\Big]\Big\}\,,   
\label{eq:bareaction}
\end{align}
where the Latin indices represent different flavor species
and $\dir$ is the covariant Dirac operator.  Denoting the vector
interaction channel term with $(V)=(\yb\gamma_{\mu}\psi)^2$ and the
axial one with $(A)=-(\yb\gamma_{\mu}\gamma_5\psi)^2$, we expect the
transition to be triggered by  the $(V)+(A)$ term which is equivalent to 
\begin{align}
 (V)+(A)=-2[(S^{\text{N}})-(P^{\text{N}})]
\label{eq:Fierz}
\end{align}
by means of a Fierz transformation. Here, $(S^{\text{N}})$ and
$(P^{\text{N}})$ denote the scalar and pseudo scalar channels in the
space of flavor nonsinglet terms,
\begin{align}
 (S^{\text{N}})&=(\yb^a\psi^b)^2=(\yb^a\psi^b)(\yb^b\psi^a),\notag\\
 (P^{\text{N}})&=(\yb^a\gamma_5\psi^b)^2=(\yb^a\gamma_5\psi^b)(\yb^b\gamma_5\psi^a)\,.
\label{eq:SandP}
\end{align}
In fact, the structure $(S^{\text{N}})-(P^{\text{N}})$ is familiar
from the Nambu--Jona-Lasinio (NJL) model and further generic models of
chiral symmetry breaking. In such models, the onset of chiral symmetry
breaking is signaled by this channel becoming RG relevant. For
instance, in the NJL model, this onset is triggered by a choice of the
four-fermion coupling being larger than some critical value. Hence, we
concentrate in the following on the NJL channel and ignore the
$(V)-(A)$ channel for the rest of the paper. The latter is expected to
stay RG irrelevant across a possible phase transition, justifying to
approximate $\lambda_-\simeq 0$ for the purpose of detecting the onset
of symmetry breaking.

Using the projectors on the left and right chiral components
\begin{align}
 P_{\text{L}} = \frac{\mathds{1}-\gamma_5}{2}&\,,\quad P_{\text{R}} = \frac{\mathds{1}+\gamma_5}{2}\,,\quad %\\
 \mathds{1}%&
 =P_{\text{L}}+P_{\text{R}}\,,
 \label{id}
\end{align}
the NJL channel can also be written as the interaction part of the Lagrangian
\begin{align}
 \mathcal{L}_{\text{int}}(\yb, \psi) = -2\bar{\lambda}(\yb_{\text{L}}^a\psi_{\text{R}}^b)(\yb_{\text{R}}^b\psi_{\text{L}}^a)\,, \quad\bar{\lambda}=2\bar{\lambda}_+.
\label{eq:NJL}
\end{align}
Here, the subscripts $\text{L},\text{R}$ represent the chiral
projections of the Dirac fermion. By means of a Hubbard-Stratonovich
transformation, the interaction term can also be expressed in terms of
a Yukawa interaction with an auxiliary scalar field,
\begin{align}
\label{eq:interactionHStrick}
\mathcal{L}_{\text{int}}(\phi, \yb, \psi) = \yb^a[P_{\text{L}}(\phi^{\dagger})_{ab}+P_{\text{R}}\phi_{ab}]\psi^b+\frac{1}{2\bar{\lambda}}\tr(\phi^{\dagger}\phi)\,.
\end{align}
The equivalence of \Eqref{eq:interactionHStrick} with \Eqref{eq:NJL}
becomes obvious with the help of the equations of motion for the chiral matrix fields $\phi$
and $\phi^{\dagger}$,
\begin{align}
 \phi_{ab} &= -2\bar{\lambda}\yb_{\text{R}}^b\psi_{\text{L}}^a\,,\notag\\
 (\phi^{\dagger})_{ab} &= -2\bar{\lambda}\yb_{\text{L}}^b\psi_{\text{R}}^a\,.
\label{eq:EoMphi}
\end{align}
From \Eqref{eq:interactionHStrick}, it is obvious that the Dirac
particles can acquire a mass if chiral symmetry gets broken by a
nonzero expectation value of the field $\phi_{ab}$.  The precise
breaking pattern is fixed by the nonzero components of $\langle
\phi_{ab}\rangle$ which in turn is determined by the minima of the
effective potential for $\phi$. In the following, we assume a diagonal
breaking pattern, $\phi_{ab} = \phi_0 \delta_{ab}$ with constant order
parameter $\phi_0$, which for $|\phi_0|>0$ breaks the chiral group
down to a residual vector symmetry familiar from QCD-like theories. In
the form of \Eqref{eq:interactionHStrick} read together with the
fermion kinetic term, we can integrate out the fermionic degrees of
freedom and obtain the standard mean-field expression for the
effective potential of the order parameter
\begin{align}
 U(\phi_0) &= \frac{\Nf}{2\bar{\lambda}}\phi_0^2 - \Nf \log \det (\dir+\phi_0)\notag\\
    &=\frac{\Nf}{2\bar{\lambda}}\phi_0^2 - \frac{\Nf}{2} \Tr\log (-\dir^2+\phi_0^2)\,,
    \label{eq:UMF}
\end{align}
where we have made use of $\gamma_5$-hermiticity of the Dirac operator
in the last step. Since we are considering a homogeneous order
parameter, the trace (as well as $\log\det$) is understood to be
already normalized by a spacetime volume factor, such that we are
considering local quantities throughout the paper.  Using the
Schwinger propertime representation, we write
\begin{align}
\label{eq:potentialWithTrace}
 U(\phi) = \frac{\Nf}{2\bar{\lambda}}\phi_0^2 + \frac{\Nf}{2}\int_0^{\infty}\frac{dT}{T}\; e^{-\phi_0^2T}\Tr\, e^{\dir^2T}\,.
\end{align}
where we encounter the trace of the heat kernel on the manifold under consideration,
\begin{align*}
 \Tr\, e^{\dir^2 T} = \Tr\, K(x, x'; T) =: K_T\,.
\end{align*}
The heat kernel $K(x, x'; T)$ satisfies
\begin{equation}
\frac{\partial}{\partial T} K=\dir^2 K, \quad \lim_{T\to 0^+}K(x,x';T) = \frac{\delta(x-x')}{\sqrt{g}}.
\end{equation}
In our analysis, the information about the nature of spacetime enters
through the trace of the heat kernel of the (squared) Dirac
operator. As this trace parametrizes the contributions of fermionic
fluctuations on all scales, the explicit evaluation of
\Eqref{eq:potentialWithTrace} would contain information about both the
local and global structure of spacetime. 

Though the propertime integration has been introduced as an auxiliary
representation, the integrand can be interpreted as the result of a
diffusion process of a fictitious particle on the spacetime within
propagation time $T$ \cite{Schwinger:1951nm,Lauscher:2005qz}. The trace enforces that the diffusion
path is closed. For a finite propertime $T$, the fictitious particle
traces out a closed path in spacetime which is localized around a
point $x$ under consideration. This path can be considered as the
spacetime path of a virtual fermionic fluctuation; this perspective
can also be made explicit by introducing a Feynman path-integral
representation of the heat kernel (worldline formalism) \cite{Feynman:1950ir,Halpern:1977ru,Schubert:2001he,Gies:2001zp,Gies:2001tj}. For
instance, the mean average distance of the diffusing particle from its
center of mass in flat space is $d=\sqrt{T/6}$ \cite{Gies:2003cv}, indicating
that $\sqrt{T}$ can be considered as a typical length scale of the
fluctuations at a fixed value of $T$.

Aiming at a statement about spacetime in the (trans-)Planckian regime,
we do not want to make an assumption about its global properties, but
intend to consider only local patches of spacetime. This is possible
by means of an RG-type analysis of \Eqref{eq:potentialWithTrace}.  For
this, we introduce a propertime regulator function $f_k$ inside the
propertime integral \cite{Liao:1994fp,Liao:1995nm},
\begin{align}
 f_k = e^{-(k^2T)^p}\,.
\label{eq:fk}
\end{align}
Here, the power $p>0$ is a parameter specifying the details of the
regularization and $k$ corresponds to an IR momentum space
regularization scale. For instance, for $p\to\infty$, all long-range
contributions for length scales $\sqrt{T}>1/k$ are cut off
sharply. For finite $p$, the length scale $1/k$ becomes a smooth
long-range cutoff. The case $p=1$ is special as it corresponds
precisely to a Callan-Symanzik regularization scheme. In the limit
$k\to0$, the insertion factor becomes $f_{k\to 0}=1$ and the
regularization is removed. Starting from the bare potential
$U_\Lambda$ at a high momentum scale $k=\Lambda$, the potential at any
IR scale $\kir$ can be constructed from
\begin{align}
\label{infraredPotential}
 U_{\kir} &= U_{\Lambda}-\int_{\kir}^{\Lambda}dk\;\partial_k U_k\,,\quad U_\Lambda=\frac{\Nf}{2\bar{\lambda}_\Lambda}\phi_0^2, \quad \bar{\lambda}_\Lambda\equiv \bar{\lambda},
\end{align}
once the RG flow of the potential is known,
\begin{equation}
\partial_k U_k =\frac{\Nf}{2}\int_0^{\infty}\frac{dT}{T}\; e^{-\phi_0^2T} \partial_k f_k K_T\,.
\label{eq:Ukflow}
\end{equation}
In \Eqref{infraredPotential}, we explicitly appended the subscript
$\Lambda$ to the bare coupling $\bar\lambda$ in order to highlight
that the bare coupling has to be fixed at the high scale in order to
define the model.

Since $\partial_k f_k\sim T^p$ for small $T$, also the short-range
fluctuations are suppressed in \Eqref{eq:Ukflow}, such that the
consequences of the fermionic fluctuations can be studied in a
Kadanoff-Wilson-spirit length scale by length scale. The evaluation of
one RG step $\sim \partial_k U_k$, typically receives contributions
from length scales $\sqrt{T}\sim 1/k$. This implies that we do not
have to know the global structure of the spacetime, but our assumptions
about the spacetime properties need to hold only over these covariant
length scales. More specifically, we assume below that the spacetime
can locally be approximated as maximally symmetric.

Though the analysis of the chiral interactions leading to
\Eqref{eq:bareaction} has been performed in $D=4$ dimensional
spacetime, the analysis of the flow of the order-parameter potential
in \Eqref{eq:Ukflow} can be performed in any $D$, though the relation
to the symmetry-breaking channel can be more involved or not
necessarily be unique in other dimensions, see \cite{Gehring:2015vja} for
an analysis in $D=3$. In higher dimensions, the perturbative
non-renormalizability of Yukawa theories suggests that more relevant
operators appear near the Gau\ss{}ian fixed point. The corresponding
regularization of UV divergencies may require higher values of $p$ for
a stronger suppression of UV modes. Independently of these technical
complications, our analysis can in principle be performed in any
dimension.

\subsection{$D=3$}

Let us begin with an analysis of the RG flow of the potential for the
case of $D=3$ spacetime dimensions. This case is highly instructive
from the viewpoint of the method: it can be treated analytically in
all detail, and does not involve further relevant operators. Since
gravitational catalysis can occur for negative curvature, we consider
spacetimes that can locally be approximated by a hyperbolic space for
Euclidean signature, corresponding to AdS spacetime for a Lorentzian
signature. The analysis could similarly be performed for spacetimes
with negative curvature in the purely spatial part with quantitatively
rather similar results \cite{Gorbar:2007kd,Gies:2013dca}. In $D=3$,
the trace of the heat kernel reads
\cite{Camporesi:1992tm,Camporesi:1995fb}
\begin{align}
 K_T = \frac{1}{8\pi^{\frac{3}{2}}T^{\frac{3}{2}}}\Big(1+\frac{1}{2}\kappa^2T\Big)\,,
\end{align}
where 
\begin{equation}
\kappa^2 = -\frac{R}{D(D-1)} = -\frac{R}{6}\geq0,
\label{eq:kappa}
\end{equation}
denotes the local curvature parameter related to the Ricci scalar
$R$. Further details of the heat kernels relevant for this work are
briefly reviewed in the appendix.

Including the propertime regularization, this leads to an effective, scale-dependent potential of the following form:
\begin{align}
 U_k &= \frac{\Nf}{2\bar{\lambda}}\phi_0^2 + \frac{\Nf}{2(4\pi)^{\frac{3}{2}}}\int_0^{\infty}\frac{dT}{T^{\frac{5}{2}}}\; e^{-\phi_0^2T}f_k \Big(1+\frac{1}{2}\kappa^2T\Big)\,.
\label{eq:UkD=3}
\end{align}
In $D=3$, the Callan-Symanzik regulator is known to be sufficient to
control the RG flow of our model. Thus, let us first choose the exponent $p=1$
for $f_k$ for simplicity; the result for general $p$ will be given below. The regularized flow of the potential with
respect to the scale $k$ then reads
\begin{align}
 \partial_kU_k(\phi) =& -\frac{2k\Nf}{2(4\pi)^{\frac{3}{2}}}\bigg[\int_0^{\infty}\frac{dT}{T^{\frac{3}{2}}}\; e^{-k^2T}\Big(e^{-\phi_0^2T}-1\Big)\label{eq:flowUkD=3}\\
  &+\frac{\kappa^2}{2}\int_0^{\infty}\frac{dT}{T^{\frac{1}{2}}}\;e^{-k^2T} \Big(e^{-\phi_0^2T}-1\Big)\bigg]\notag\\
  =\,\,\,\,\,\,\,\,&\!\!\!\!\!\!\!\!\frac{\Nf}{4\pi}\bigg[k^2\Big(\sqrt{1+\frac{\phi_0^2}{k^2}}-1\Big)-\frac{\kappa^2}{4}\Big(\frac{k}{\sqrt{\phi_0^2+k^2}}-1\Big)\bigg].\notag 
\end{align}
Upon insertion into \Eqref{infraredPotential}, the effective potential at the scale $\kir$ can be computed, yielding
\begin{align}
 U_{\kir} =& -\frac{\Nf}{2}\phi_0^2\Big(\frac{1}{\blamc}-\frac{1}{\blamL}-\frac{\kir}{4\pi}\Big)\notag\\
    &+\frac{\Nf}{12\pi}\Big((\phi_0^2+\kir^2)^{\frac{3}{2}}-\frac{3}{2}\kir\phi_0^2-\kir^3\Big)\notag\\
    &-\frac{\Nf}{16\pi}\kappa^2\Big(\sqrt{\phi_0^2+\kir^2}-\kir\Big)\,,
 \label{dimensionalPotential3dim}
\end{align}
where we have introduced the (scheme-dependent) critical coupling
$\bar{\lambda}_{\text{cr}}=4\pi/\Lambda$, and dropped terms of order
$\mathcal{O}(1/\Lambda)$.

The physics described by this effective potential can be read off line
by line: the first line describes the mass-like term in the
potential. For subcritical coupling $\blamL<\blamc$, the mass-like
term remains positive for any $\kir$ implying that the system in flat
space remains in the symmetric phase with a minimum $\phi_0=0$ and
does not develop fermion masses. For supercritical couplings
$\blamL>\blamc$, the mass-like term becomes negative below a certain
critical IR scale $\kir$, indicating that the potential develops a
nontrivial minimum $\phi_0^2>0$. The system hence exhibits chiral
symmetry breaking and fermion mass generation already in flat
space. The second line does not contribute to the mass-like term $\sim
\phi_0^2$ upon Taylor expansion. For large $\phi_0$ it grows $\sim +
\phi_0^3$, ensuring stability of the potential. The third line
represents the contribution due to nonzero curvature, being manifestly
negative. In the limit $\kir\to0$, it is linear in the field $\phi_0$
and thus dominates for small field amplitudes. In this way, it induces
a nonzero $\phi_0$ and inevitably drives the system to chiral symmetry
breaking and fermion mass generation -- the essence of gravitational
catalysis \cite{Buchbinder:1989fz,Buchbinder:1989ah,Buchbinder:1989ah,Inagaki:1993ya,Sachs:1993ss,Elizalde:1995kg,Kanemura:1995sx,Inagaki:1995bk,Inagaki:1996nb,Geyer:1996wg,Geyer:1996kg,Miele:1996rp,Vitale:1998wm,Inagaki:1997kz,Hashida:1999wb,Gorbar:2007kd,Hayashi:2008bm,Inagaki:2010py,Sasagawa:2012mn,Gorbar:1999wa,Ebert:2008pc,Gorbar:2008sp,Gies:2013dca}.

However, gravitational catalysis receives its relevant contributions
from the deep IR, i.e., the long-wavelength modes. In order to
dominate the mass spectrum, the curvature has to be such that the
hyperbolic space is an adequate description also on large length
scales. Within our RG description, we make the less severe assumption
that the hyperbolic space is an adequate description only up to
lengths scales of order $1/\kir$. Whether or not the potential
develops a nonzero minimum then is decided by the competition between
the first and the third line of \Eqref{dimensionalPotential3dim}.

Since we are interested in curvature-induced symmetry breaking, we
assume that the fermionic interactions are subcritical,
$\blamL\leq\blamc$, such that the mass-like term in the first line is
bounded from below by
\begin{equation}
-\frac{\Nf}{2}\phi_0^2\Big(\frac{1}{\blamc}-\frac{1}{\blamL}-\frac{\kir}{4\pi}\Big) \geq\frac{\Nf\kir}{8\pi} \phi_0^2.
\label{eq:D=3res1}
\end{equation}
The only other term contributing to the mass-like term arises from the
curvature-dependent third line of \Eqref{dimensionalPotential3dim}:
\begin{equation}
-\frac{\Nf}{16\pi}\kappa^2\Big(\sqrt{\phi_0^2+\kir^2}-\kir\Big) =
- \frac{\Nf\kir}{32\pi} \frac{\kappa^2}{\kir^2} \phi_0^2 + \mathcal{O}(\phi_0^4)
\label{eq:D=3res2}
\end{equation}
Comparing the last two equations tells us that gravitational catalysis does not induce chiral symmetry breaking and fermion mass generation as long as hyperbolic-curvature parameter satisfies
\begin{equation}
\frac{\kappa^2}{\kir^2}\leq4.
\label{eq:kappabound3D}
\end{equation}
In terms of the negative scalar (spacetime) curvature, this implies
\begin{align}
\label{eq:bound3D}
 |R|=6\kappa^2\leq24\kir^2\,, \quad (\text{for}\,\,R<0)
\end{align}
inhibits the occurrence of the nontrivial minimum of the effective
potential and thus fermion mass generation induced by a negative
mass-like term. Equation \eqref{eq:bound3D} represents our first
example of a curvature bound from gravitational catalysis: in line
with our assumptions we conclude that a fermionic particle-physics
system will not be plagued by curvature-induced chiral symmetry
breaking, as long as the local curvature of spacetime patches averaged
over the scale of $1/\kir$ satisfies the bound \eqref{eq:bound3D}.

Some comments are in order: (i) from the derivation, it is obvious that
a study of the mass-like term $\sim \phi_0^2$ is sufficient to obtain
a curvature bound. Of course, the global structure of an effective
potential could be such that a nontrivial minimum exists even for a
positive mass-like term. In that case, the true curvature bound would
even be stronger than the one derived from the mass-like term. (In the
present $D=3$ dimensional system, this does not happen at mean-field
level.).

(ii) The curvature bound is independent of the self-couplings, because
of our estimate performed in \Eqref{eq:D=3res1}. The equal sign holds
for bare couplings exactly tuned to criticality, i.e., the maximum
value of the self-coupling that does not lead to chiral symmetry
breaking in the IR. Therefore, the bound limits the curvature and
coupling regime where the system is safe from fermion mass generation
through gravitational catalysis. Whether or not fermion mass
generation sets in if the bound is violated depends on further details
of the system such as the fermion couplings. 

(iii) The bound is naively scheme-dependent in the sense that the
prefactor (24 in the present case) depends on the way the fluctuation
averaging procedure is performed. In the calculation so far, we used a
Callan-Symanzik regulator that suppresses long-wavelength modes beyond
the scale $1/\kir$ exponentially. In fact, the calculation can
straightforwardly be performed for the general regulator
\eqref{eq:fk}. For general $p$, we obtain
\begin{equation}
\frac{\kappa^2}{\kir^2}\leq \frac{2 \Gamma(1-{\scriptstyle \frac{1}{2p}})}{\Gamma(1+{\scriptstyle \frac{1}{2p}})}.
\end{equation}
For $p=1$, we obtain \eqref{eq:kappabound3D} and \eqref{eq:bound3D}
again, whereas we find in the sharp cutoff limit $p\to\infty$
\begin{align}
\label{eq:bound3Dpinfty}
 |R|\leq12\kir^2\,, \quad (\text{for}\,\,p\to \infty,R<0).
\end{align}
Comparing this to \eqref{eq:bound3D}, the curvature bound naively
seems to be stronger for $p\to\infty$. However, this simply reflects
the fact that the length scale of the fluctuations $1/\kir$ is
effectively shorter for the sharp cutoff than for the smooth
exponential regulator, where the fluctuations extending even further
out are only suppressed but not cut off. Hence, it is plausible to say
that $\kir|_{p\to\infty}$ is effectively larger than
$\kir|_{p=1}$. This goes hand in hand with the inversely behaving
prefactor. We consider this as an indication that the curvature bound
itself has a scheme-independent meaning: the scheme-dependence of the
prefactors in the bound should be viewed as a parametrization of the
fluctuation averaging process that has to be matched with the
procedure that determines the averaged curvature.

\section{$D=4$ dimensional spacetime}
\label{sec:fourDim}

Let us now turn to the physically more relevant case of
$D=4$ dimensional spacetime. The analysis is conceptually complicated
by the appearance of two more relevant operators coming along with
physical couplings, and technically more involved because of the
structure of the heat kernel. Nevertheless, it is possible to capture
the essential behavior analytically by making use asymptotic
heat-kernel expansions and a simple interpolation. The full result
is, of course, analyzed below by straightforward numerical
integration. We start with the representation of the heat-kernel trace
as a one-parameter integral \cite{Camporesi:1992tm,Camporesi:1995fb}
\begin{align}
 K_T =& \frac{2}{(4\pi T)^2}\int_0^{\infty} du\;e^{-u^2}u(u^2+\kappa^2T)\coth(\frac{\pi u}{\kappa\sqrt{T}})\,.
\label{eq:KTD=4}
\end{align}
Using the asymptotic expansions of the $\coth$ function, cf. Eqs.~\eqref{cothSmallT} and \eqref{cothLargeT}, the weak and strong curvature expansions of the heat kernel read
\begin{align}
 K_T =& \frac{2}{(4\pi T)^2}(1+\kappa^2T+\dots),\quad \kappa^2T\ll 1,\label{eq:kTD4exp1}\\
 K_T =& \frac{1}{(4\pi T)^2}\Big(\frac{\kappa^3T^{\frac{3}{2}}}{\pi} + \frac{3+\pi^2}{6\sqrt{\pi}}\kappa T^{\frac{1}{2}}+\dots\Big),\quad\kappa^2T\gg 1\,.\label{eq:kTD4exp2}
\end{align}
For a simple qualitative, still asymptotically exact estimate, we use an interpolating approximation of the heat kernel that allows for a fully analytical treatment,
\begin{equation}
 K_T \simeq \frac{2}{(4\pi T)^2}(1+\kappa^2T+\frac{\kappa^3T^{\frac{3}{2}}}{\pi}  ). \label{eq:kTD4exp3}
\end{equation}
Upon insertion of the heat kernel into Eqs.~\eqref{infraredPotential}
and \eqref{eq:Ukflow}, a first difference to the $D=3$ case is the
occurrence of a logarithmic UV divergence of the type $\sim \phi_0^4
\ln \Lambda$. This is expected as $\phi^4$ is a marginal operator in
$D=4$, the coupling of which corresponds to a new and independent
physical parameter. The proper definition of the particle system
requires to also define an initial condition for the flow of this
operator, i.e., to put a counter-term at the high scale
$\Lambda$. This is then fixed by demanding for a specific physical
renormalized value for the $\phi^4$ coupling in a long-range
experiment.

For our purposes, these details are, in fact, not relevant, as the
$\phi^4$ coupling cannot inhibit chiral symmetry breaking. Once, the
mass-like term $\sim\phi_0^2$ triggers the onset of a chiral
condensate, the $\phi^4$ coupling will take influence on the final
value of the condensate $\phi_0$; this is, however, irrelevant for the
curvature bound. For consistency, we only assume that the renormalized
$\phi^4$ coupling is such that the potential is stable towards large
fields.

As we have seen in the $D=3$ case, we can obtain a curvature bound by
solely studying the $\phi_0^2$ term of the potential. Using the
approximate form of the heat kernel \eqref{eq:kTD4exp3}, we obtain the analytic estimate to this order:
\begin{align}
 U_{\kir}\Big|_{\phi_0^2} 
 =&-\frac{\Nf\phi_0^2}{2}\Big[\frac{1}{\blamc}-\frac{1}{\bar{\lambda}_{\Lambda}}-\Gamma\Big(1-\frac{1}{p}\Big)\frac{\kir^2}{(4\pi)^2}\notag\\
 &+2\frac{\kappa^3}{\kir}\frac{\Gamma\Big(1+\frac{1}{2p}\Big)}{\sqrt{\pi}}\Big]-\frac{\Nf\phi_0^2}{(4\pi)^2}\kappa^2\log\Big(\frac{\Lambda}{\kir}\Big)\,,
\label{potentialIntegrated4DApprox}
\end{align}
again dropping terms of order $\mathcal{O}(1/\Lambda)$.  As before,
the diverging contribution coming from the flat part of the heat
kernel is indicative of the critical value of the coupling constant,
\begin{align*}
 \blamc=\frac{(4\pi)^2}{\Lambda^2\Gamma\Big(1-\frac{1}{p}\Big)}\,,
\end{align*}
As a new feature in $D=4$, we observe a new logarithmically divergent
term $\sim\ln \Lambda$ in \Eqref{potentialIntegrated4DApprox}. This
term corresponds to a new, power-counting marginal operator of the
form $\phi^2 R$, which again comes along with a new physical parameter
to be fixed by renormalization. Hence, we introduce an initial
condition for this operator at the high scale with a bare coupling $\xi_\Lambda$:
\begin{equation}
\label{microMarginalOp4Dim}
 U_{\Lambda}|_{\phi^2R}=\Nf\xi_{\Lambda}\phi^2R.
\end{equation}
Upon inclusion of \Eqref{microMarginalOp4Dim}, the effective potential
at the scale $\kir$ receives an overall contribution of the form
\begin{align}
 U_{\kir}|_{\phi^2R} =& -\Big(\xi_{\Lambda}\Nf+\frac{3\Nf}{4\pi^2}\log\Big(\frac{\Lambda}{\kir}\Big)\Big)\phi_0^2|R|\notag\\
  \equiv& -\Nf\xi_{\kir}\phi_0^2|R|\,,\label{microMarginalCounterterm4D}
\end{align}
where we have made use of the relation $\kappa^2 = \frac{|R|}{12}$, $R<0$, 
in $D=4$. Here we have introduced the long-range parameter
$\xi_{\kir}$ that, in principle, has to be fixed by a physical
measurement. For our analysis, we will consider it as a free
parameter. As a consequence, the curvature bound depends
parametrically on this physical coupling. Assuming again, that the
fermion self-interactions are subcritical $\blamL\leq\blamc$, we obtain
again a bound on the curvature parameter for which no chiral symmetry
breaking occurs:
\begin{equation}
 \frac{\kappa^3}{\kir^3}+\frac{4}{3}\frac{\pi^{\frac{5}{2}}}{\Gamma\Big(1+\frac{1}{2p}\Big)}\xi_{\kir}\frac{\kappa^2}{\kir^2}\leq
	\frac{\sqrt{\pi}}{2}\frac{\Gamma\Big(1-\frac{1}{p}\Big)}{\Gamma\Big(1+\frac{1}{2p}\Big)}\,.
\label{boundApprox4dim}
\end{equation}
The divergence of the right-hand side for $p\to1$, where the bound
seems to disappear, is an artifact of the Callan-Symanzik regulator
which is insufficient to control all UV divergences in $D=4$. In order
to stay away from this artifact, we consider regulators in the range
$p\in [2,\infty]$.

For a comparison with the $D=3$ case, let us first set $\xi_{\kir}=0$ and consider limiting regulator values,
\begin{equation}
\frac{\kappa^3}{\kir^3}\Big|_{p=2}\leq\frac{\sqrt{\pi}}{2} \frac{\Gamma({\scriptstyle \frac{1}{2}})}{\Gamma({\scriptstyle \frac{5}{4}})}, \qquad \frac{\kappa^3}{\kir^3}\Big|_{p\to\infty}\leq \frac{\sqrt{\pi}}{2}.
\label{eq:approxboundD4}
\end{equation}
From \eqref{boundApprox4dim}, it is obvious that the bound gets
stronger (weaker) for positive (negative) coupling $\xi_{\kir}$. Most
importantly, there is a nontrivial bound for any finite value of
$\xi_{\kir}$.

While \Eqref{eq:approxboundD4} has been derived analytically based on
the interpolating approximation \eqref{eq:kTD4exp3} for the heat
kernel, a full calculation can be performed numerically. For this, we
first have to isolate the divergent pieces by hand, treat them
analytically as before. In fact, all divergent parts are related to
the small curvature expansion of the heat kernel, i.e., to the
expansion coefficients displayed in \Eqref{eq:kTD4exp1}. Treating them
separately as before leaves us with a triple integral over the heat
kernel parameter $u$ in \Eqref{eq:KTD=4}, the propertime $T$ and the
RG scale $k$. A transition to dimensionless integration variables
$t=\kappa^2 T$ and $\sigma=k/\kappa$ yields an integral representation
depending only on the dimensionless parameter ratio $\kappa/\kir$. The
mass-like term of the effective potential then acquires the form
\begin{align}
 U_{\kir}\Big|_{\phi_0^2} =& -\frac{\Nf\phi_0^2}{2}\Big[\frac{1}{\blamc}-\frac{1}{\bar{\lambda}_{\Lambda}}+\kappa^2\Aa\Big(\frac{\kappa}{\kir};p\Big)\Big]\notag\\
   &-12\Nf\xi_{\kir}\phi_0^2\kappa^2\,, \label{eq:UknumD4}
\end{align}
with the function $\Aa$ to be evaluated by numerical
integration. Assuming subcritical fermion interactions
$\blamL\leq\blamc$, the curvature bound can be expressed as
\begin{align}
  \frac{1}{2}\Aa\Big(\frac{\kappa}{\kir};p\Big)+12\xi_{\kir}\geq 0\,,\label{eq:boundD4num}
\end{align}
in order to avoid fermion mass generation from gravitational
catalysis. 
\begin{figure}
 \includegraphics[width=0.45\textwidth]{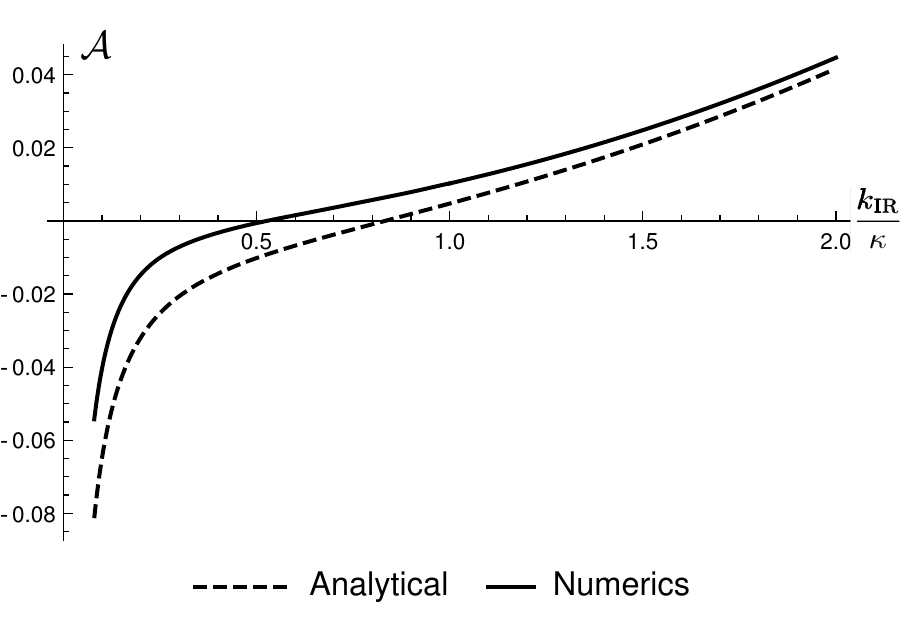}
 \caption{Scaling of the mass-like term of the effective potential: the function $\Aa$, cf. \Eqref{eq:UknumD4}, (solid line) as a function of the inverse curvature parameter $\kir/\kappa$ is compared to the analytical approximation obtained, cf. \Eqref{potentialIntegrated4DApprox}, (dashed line) for the case $p=2$. For the case of $\xi_{\kir}=0$, positive values of $\Aa$ are compatible with the absence of chiral symmetry breaking and the existence of chiral fermions at low energies. The zero crossing corresponds to the curvature bound for gravitational catalysis.}
 \label{fig:comparison}
\end{figure}
The function $\Aa$ is plotted in Fig.~\ref{fig:comparison}
as a function of $\kir/\kappa$ for $p=2$ (solid line). For comparison,
the dashed line represents the result from the analytical
interpolation matching the full behavior qualitatively for all
curvatures. The strong and weak curvature asymptotics matches very
well: we have checked that the leading powerlaws for both results are
the same with coefficients agreeing within an error below the $1\%$
level. In the intermediate curvature region, the deviations between
the numerical result and the analytical estimate are larger.

For $\xi_{\kir}=0$, the zero of the curve marks the curvature bound,
since positive values of $\Aa$ are compatible with the absence of
chiral symmetry breaking. From the numerical analysis we obtain the
curvature bound,
\begin{equation}
\frac{\kappa}{\kir}\Big|_{p=2}\leq1.8998 , \qquad \frac{\kappa}{\kir}\Big|_{p\to\infty}\leq 1.5757.
\label{eq:approxboundD4num}
\end{equation}
for the two limiting regulators,
showing that the full solutions deviate from the approximated ones by about 40\%.

\begin{figure}
 \includegraphics[width=0.45\textwidth]{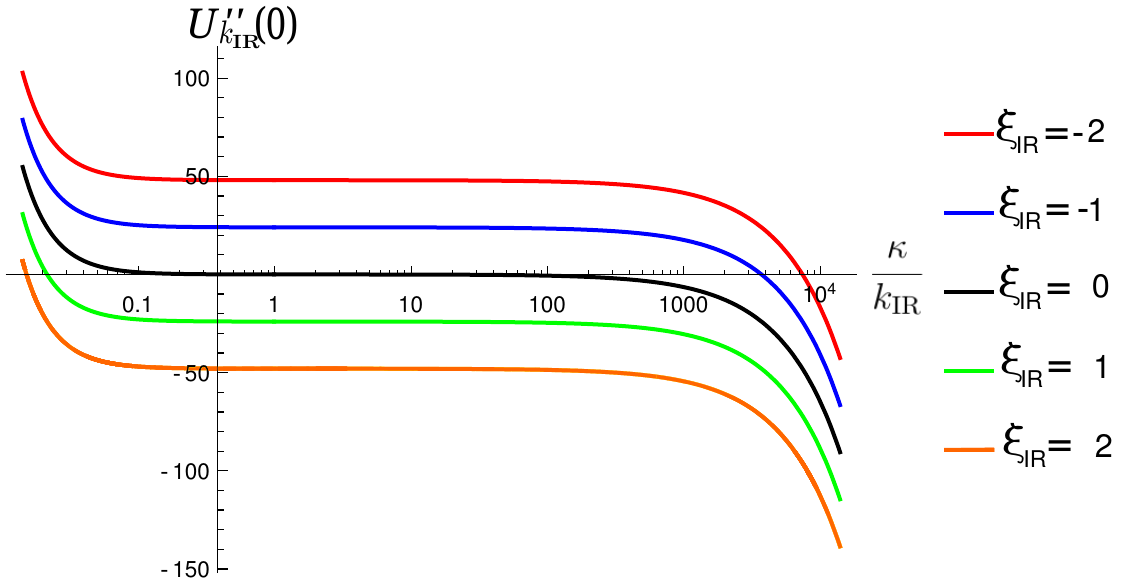}
 \caption{Scaling behavior of the curvature of the effective potential at the origin  (normalized by $\Nf\kappa^2/2$) as a function of the curvature parameter $\kappa/\kir$ for the case $p=2$ and different values of the marginal coupling $\xi_{\kir}$. Positive values are compatible with the existence of light chiral fermions. The zero crossing marks the curvature bound which is strengthened for increasing values of $\xi_{\kir}$.}
 \label{fig:scalingDifferentXi4D}
\end{figure}

A finite $\xi_{\kir}$ parameter corresponds to a linear vertical shift
of the graph in Fig.~\ref{fig:comparison} and a corresponding shift of
the zero crossing marking the curvature bound. Figure
\ref{fig:scalingDifferentXi4D} shows the curvature of the effective
potential at the origin (normalized by $\Nf\kappa^2/2$) %\colH{[H: please check]}
as a function of the curvature parameter
$\kappa/\kir$ for various values of $\xi_{\kir}$. Positive values are
compatible with the existence of light chiral fermions.

\section{Higher dimensions}
\label{sec:higherDim}

\subsection{$D = 6$}
\label{sec:dim6}

It is instructive to also study curvature bounds in higher
dimensions. Perturbative nonrenormalizability implies that further
relevant operators and thus physical couplings have to be accounted
for; still, for any finite dimension, also the number of additional
couplings is finite at mean-field level. As before, we have to pay
attention only to those operators that couple to the mass-like term in
the effective potential. Other operators do not directly take
influence on the curvature bound for chiral symmetry.

In $D=6$ dimensional spacetime, one further divergence of this type is
encountered requiring to consider one more physical parameter. As
before, the divergences are in correspondence with the
small-propertime expansion of the heat kernel for which we need to
retain only the $0$-th order of the hyperbolic cotangent expansion
inside the heat kernel,
\begin{align}
 K_T^{\text{div}}=&\frac{1}{(4\pi T)^3}\int_0^{\infty}du\;e^{-u^2}u(u^2+\kappa^2T)(u^2+4\kappa^2T)\notag\\
 =&\frac{1}{(4\pi T)^3}(1+\frac{5}{2}\kappa^2T+2\kappa^4T^2)\,.
 \label{6dimDivergencies}
\end{align}
The divergencies associated with the curvature-dependent terms are
controlled by initial conditions for the two operators 
\begin{align}
 U_\Lambda|_{\phi^2R, \phi^2R^2}=\Nf\xi_{\Lambda}\phi_0^2R+\Nf\chi_{\Lambda}\phi_0^2R^2\,.
\label{eq:ctD6}
\end{align}
Adding these two operators to the terms arising from
\Eqref{6dimDivergencies}, yields the following contributions to the
mass-like term in the effective potential:
\begin{align}
 U_{\kir}^{\text{div}}\Big|_{\phi_0^2} =& -\frac{\Nf\phi_0^2}{2}\Big\{\frac{\Lambda^4-\kir^4}{2(4\pi)^3}\Gamma\Big(1-\frac{2}{p}\Big)\notag\\
    &+\kappa^2\Big[60\xi_{\Lambda}+5\frac{\Lambda^2-\kir^2}{2(4\pi)^3}\Gamma\Big(1-\frac{1}{p}\Big)\Big]\notag\\
    &+\kappa^4\Big[-1800\chi_{\Lambda}+\frac{4}{(4\pi)^3}\log\Big(\frac{\Lambda}{\kir}\Big)\Big]\Big\}\notag\\
 \equiv&-\frac{\Nf\phi_0^2}{2}\Big\{ \frac{1}{\blamc}-\frac{\kir^4}{(4\pi)^3}\Gamma\Big(1-\frac{2}{p}\Big) \notag\\
 & -2\xi_{\kir}R-2\chi_{\kir}R^2\Big\}\,. \label{eq:UkdivD6}
\end{align}
Here, we have used that $\kappa^2=\frac{|R|}{D(D-1)}=\frac{|R|}{30}$ in $D=6$, and identified the critical coupling
\begin{align}
 \blamc=\frac{2(4\pi)^3}{\Lambda^4\Gamma\Big(1-\frac{2}{p}\Big)}\,. \label{eq:critcoupD6}
\end{align}
The parameter $\xi_{\Lambda}$ has positive mass dimensions
($[\xi_{\Lambda}]=2$) and thus the operator $\phi_0^2R$ is now a
power-counting relevant operator, while $\phi_0^2R^2$ is marginal and
the corresponding coupling $\chi_{\Lambda}$ has vanishing mass
dimensions. The curvature-dependent terms in the last line of
\Eqref{eq:UkdivD6} are finite and need to be fixed by a
measurement. As before, the divergence hidden in the critical coupling
will be balanced by the initial condition for the bare coupling
$\blamL$.

\begin{figure}
 \includegraphics[width=0.45\textwidth]{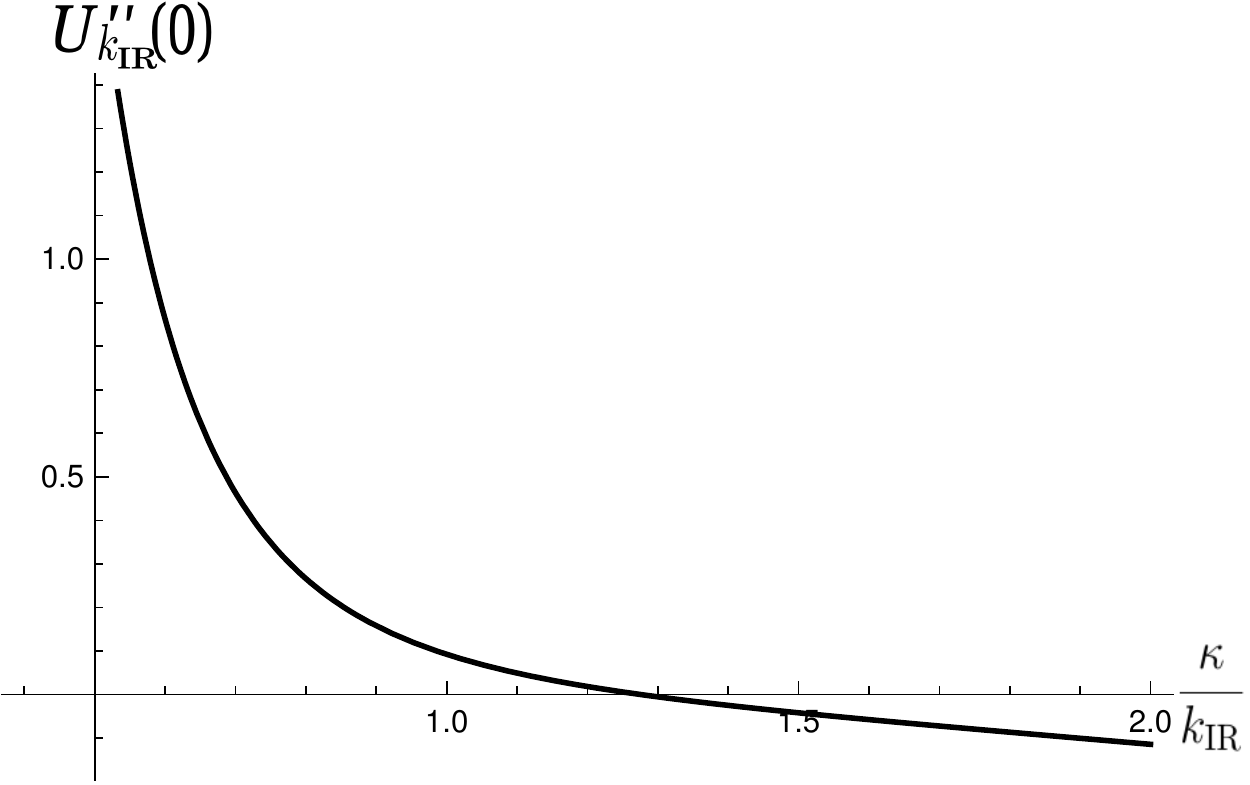}
 \caption{$D=6$ scaling behavior of the curvature of the effective
   potential at the origin (normalized by $\Nf\kappa^2/2$) as a
   function of the curvature parameter $\kappa/\kir$ for the case
   $p=4$ and $\xi_{\kir}=0=\chi_{\kir}$. Positive values are
   compatible with the existence of light chiral fermions. The zero
   crossing marks the curvature bound.}
 \label{fig:scalingSixDim}
\end{figure}

This concludes the analytical treatment of the divergent parts. The
remaining regular part of the effective potential can then be
integrated straightforwardly by numerical means as in the $D=4$
case. In order to stay away from regulator artifacts, we choose the
regulator parameter in the range $p\in[4,\infty]$. With the usual
assumption of subcriticality, the dependence of the resulting
mass-like term of the effective potential (normalized by
$\Nf\kappa^2/2$) as a function of the curvature parameter
$\kappa/\kir$ for the case $p=4$ and all further couplings
$\xi_{\kir}, \chi_{\kir}$ set to zero is depicted in
Fig.~\ref{fig:scalingSixDim}.

For a fair comparison of the curvature bounds for different spacetime dimensions, two conditions need to be met: (1) the physical parameters have to be chosen such that the relevant operator content is comparable, (2) the same $p$ parameter needs to be employed for the regularization procedure.

For the first condition, we simply set all independent scalar-curvature couplings to zero, $\xi_{\kir}=0=\chi_{\kir}$. For the second condition, we first check $p=4$ for numerical simplicity. This results in 
\begin{align}
 &\frac{\kappa}{\kir}\leq1.7039\,,\quad D=4\,, \quad\xi_{\kir}=0\,,\label{eq:boundcomp}\\
 &\frac{\kappa}{\kir}\leq1.2763\,, \quad D=6\,, \quad\xi_{\kir}=0\,,\quad\chi_{\kir}=0\,.\notag 
\end{align}
The same analysis can be performed in the $p\to\infty$ limit. This scenario can be implemented noticing that the cutoff function reduces to a Heaviside $\theta$ function centered in $T=\frac{1}{k^2}$ and its derivative is therefore a Dirac $\delta$ distribution.
In six dimensions, we obtain
\begin{align}
 \label{bound6pInfinite}
 \frac{\kappa}{\kir}\Big|_{p\to\infty}\leq1.0561\,.
\end{align}
We observe that, for both values of $p$, the bound for the dimensionless curvature parameter
decreases with increasing the spacetime dimensions (compare with \Eqref{eq:approxboundD4num}). We verify this
circumstantial evidence in the next section for all odd dimensions. A
general discussion follows below.

\subsection{Odd dimensions: $D = 2n+1$}
\label{sec:generalOdd}

The odd dimensional case is more easily analytically accessible thanks
to the absence of the hyperbolic cotangent in the heat kernel
(cf. \eqref{generalHKodd} and \eqref{generalHKeven}). In line with the
preceding studies, we associate the curvature bound with a possible
sign change of the mass-like term in the effective potential. Thus, it
suffices to focus on the $\phi_0^2$ order of the effective
potential. Inserting \eqref{generalHKodd} into \eqref{infraredPotential}
and expanding in powers of the field, we obtain
\begin{align}
 U_{\kir}\Big|_{\phi_0^2}=& U_{\Lambda}\Big|_{\phi_0^2}\frac{\Nf\phi_0^2}{2}\Bigg[-\frac{4p \kappa^{D-2}}{(4\pi)^{\frac{D}{2}}\Gamma\Big(\frac{D}{2}\Big)}\times\notag\\
    &\times\int_{\frac{\kir}{\kappa}}^{\frac{\Lambda}{\kappa}}d\sigma\;\int_0^{\infty}dt\;t^{p-\frac{D}{2}}\sigma^{2p-1}e^{-(\sigma^2 t)^p}\times\notag\\
    &\times\int_0^{\infty}du\;e^{-u^2}\prod_{j=\frac{1}{2}}^{\frac{D}{2}-1}(u^2+j^2t)\Bigg]\,,\label{eq:UkD2np1}
\end{align}
where we have defined the dimensionless integration variable as
$\sigma = \frac{k}{\kappa}$ and $t = \kappa^2 T$. The effective potential can be decomposed into the following building blocks,
\begin{equation}
 U_{\kir}=U_{\Lambda}\Big|_{\kappa=0} + U_{\kir}\Big|_{\kappa=0} + U_{\kir,\kappa}^{\text{reg}} +U_{\Lambda}\big|_{\phi^2R^n} +U_{\kir}\big|_{\phi^2R^n},
\label{eq:Ukdecomp}
\end{equation}
which we discuss separately in the following, concentrating on the
mass-like term $\sim\phi_0^2$. The first two terms correspond to the
contribution which is present in flat space. By renormalizing the
fermionic self-interaction, these terms exhibit the balance between
the bare coupling $\blamL$ and the leading cutoff divergence. The
latter arises from the monomial containing the highest power of $u$ in
the product in \Eqref{eq:UkD2np1} $\sim u^{D-1}$, and can be
summarized in the definition of the critical coupling
\begin{align}
	\blamc=\frac{(4\pi)^{\frac{D}{2}}(D-2)}{2\Lambda^{D-2}\Gamma\Big(1-\frac{D}{2p}+\frac{1}{p}\Big)}\,.
\label{eq:critcoupD}
\end{align}
For the flat-space part, we thus obtain
\begin{align}
&U_{\Lambda}\Big|_{\phi_0^2,\kappa=0} + U_{\kir}\Big|_{\phi_0^2,\kappa=0} \notag\\
&\quad = \frac{\Nf\phi_0^2}{2} \left[ \frac{1}{\blamL}-\frac{1}{\blamc} + 
\frac{2\Gamma\Big(1-\frac{D}{2p}+\frac{1}{p}\Big)}{(4\pi)^{\frac{D}{2}}(D-2)}\kir^{D-2}\right].
\label{eq:flatD}
\end{align}
The only a priori UV-regular term in \Eqref{eq:UkD2np1} comes from the $u$-independent monomial arising 
from the product inside the last integral. It contains the relevant curvature dependence for gravitational catalysis:
\begin{align}
 U_{\kir,\kappa}^{\text{reg}}\Big|_{\phi_0^2} =& -\frac{\Nf\phi_0^2}{2}\frac{4p \kappa^{D-2}}{(4\pi)^{\frac{D}{2}}\Gamma\Big(\frac{D}{2}\Big)}\notag\\
    &\times\int_{\frac{\kir}{\kappa}}^{\frac{\Lambda}{\kappa}}d\sigma\;\int_0^{\infty}dt\;t^{p-\frac{D}{2}}\sigma^{2p-1}e^{-(\sigma^2 t)^p}\notag\\
    &\times\int_0^{\infty}du\;e^{-u^2}\frac{\Gamma^2\Big(\frac{D}{2}\Big)}{\pi}t^{\frac{D-1}{2}}\notag\\
=& -\frac{\Nf\phi_0^2}{2}\frac{2\Gamma\Big(\frac{D}{2}\Big)\Gamma\Big(1+\frac{1}{2p}\Big)}{(4\pi)^{\frac{D}{2}}\sqrt{\pi}} \frac{\kappa^{D-1}}{\kir}\,,
\label{regularOpOddDim}
\end{align}
where we have taken the limit $\Lambda\to\infty$ in the last line.

All other monomials in the product of \Eqref{eq:UkD2np1} carry UV
divergencies, thus indicating the necessity to provide initial
conditions for further operators. In total, we need $\frac{D-3}{2}$
operators with scalar-curvature couplings and correspondingly many
physical parameters to be fixed by a measurement. 
The required
operators are of the form $\Nf\xi_{\Lambda,
    m}\phi^2R^{n}$. Here, we choose conventions such that the index $m$
corresponds to a specific monomial in the above expression and
$\xi_{\Lambda, m}$ parametrizing the initial condition for the bare
coupling to be fixed. In order to analyze these contributions, we
represent the polynomial part of the heat kernel as
\begin{align}
 \prod_{j=\frac{1}{2}}^{\frac{D}{2}-1}(u^2+j^2t) = \sum_{m=0}^{\frac{D-1}{2}}C_m u^{2m}t^{\frac{D-1}{2}-m}\,,
\label{eq:polynomial}
\end{align}
where $C_m$ denotes the numerical coefficients arising from the
product. The resulting curvature dependence for each $m$ then results
in a power $R^n$ with $n=\frac{D-1}{2}-m$.  The $m=0$ term corresponds
to the regular monomial computed in \eqref{regularOpOddDim}, while the
$m=\frac{D-1}{2}$ term equals the curvature-independent part of the
heat kernel, already dealt with in \Eqref{eq:flatD}.  The remaining
terms with $1\leq m\leq (D-3)/2$ in combination with the additional
bare scalar-curvature operators thus make up for the last two terms in
our decomposition \eqref{eq:Ukdecomp} of the effective potential, yielding
\begin{align}
U_{\Lambda}\big|_{\phi_0^2R^n} &+U_{\kir}\big|_{\phi_0^2R^n} = -\Nf\phi_0^2\sum_{m=1}^{\frac{D-3}{2}}\kappa^{D-1-2m}\notag\\
 &\times\bigg\{(-1)^{\frac{D-1}{2}-m-1}\xi_{\Lambda, m}[D(D-1)]^{\frac{D-1}{2}-m}\notag\\
 &+\frac{C_m\Gamma\Big(m+\frac{1}{2}\Big)\Gamma\Big(1-\frac{m}{p}+\frac{1}{2p}\Big)}{(4\pi)^{\frac{D}{2}}\Gamma\Big(\frac{D}{2}\Big)(2m-1)}\notag\\
 &\times(\Lambda^{2m-1}-\kir^{2m-1})\bigg\}\notag\\ 
 \equiv&-\Nf\phi_0^2\sum_{m=1}^{\frac{D-3}{2}}(-1)^{\frac{D-1}{2}-m-1}\kappa^{D-1-2m}\xi_{\kir,m},
\label{divergentOpOddDim}
\end{align}
As before, the $\Lambda$-dependent terms combine with the bare
couplings such that the long-range couplings $\xi_{\kir,m}$ are
formed; for a physical system, the latter are finite and have to be
fixed by a measurement. It is clear that possible curvature bounds
will depend on these couplings. For the reason of comparing theories
with different dimensionality, we set all these couplings to zero
$\xi_{\kir,m}=0$ at the scale $\kir$.  Let us study two cases
explicitly.

\subsubsection{$D=5$}

Inserting Eqs.~\eqref{eq:flatD} and \eqref{regularOpOddDim} into \Eqref{eq:Ukdecomp} and using that \Eqref{divergentOpOddDim} gives a vanishing contribution for $\xi_{\kir,m}=0$, it is
straightforward to obtain the following result for the mass-like term
of the scale-dependent effective potential in $D=5$ dimensional
spacetime:
\begin{align*}
  U_{\kir}^{\text{D=5}}\Big|_{\phi_0^2} =&\frac{\Nf\phi_0^2}{2}\Big[\frac{1}{\blamL}-\frac{1}{\blamc}+\frac{\Gamma\big(1-\frac{3}{2p}\big)}{48\pi^{\frac{5}{2}}}\kir^3\\
    &
    -\frac{3\Gamma\big(1+\frac{1}{2p}\big)}{64\pi^{\frac{5}{2}}}\frac{\kappa^{4}}{\kir}\Big]\,. 
\end{align*}
Assuming again a subcritical coupling as initial condition of the
flow, we can identify the bound for the ratio between the curvature
parameter and the averaging scale, below which symmetry breaking is not
catalyzed gravitationally:
\begin{align}
\Big(\frac{\kappa}{\kir}\Big)^4\leq& \frac{4}{9}\frac{\Gamma\Big(1-\frac{3}{2p}\Big)}{\Gamma\Big(1+\frac{1}{2p}\Big)}\,.
        \label{eq:boundD5}
\end{align}
In order to stay away from artifacts arising from insufficient
regulators, we choose $p$ in the interval $p\in[2, \infty]$. For the
two extremal cases, we have:
\begin{align}
 &\frac{\kappa}{\kir}\leq\frac{2}{\sqrt{3}} \simeq 1.154 \quad\text{for} \quad p=2\,,\label{eq:boundD5p2}\\
 &\frac{\kappa}{\kir}\leq\sqrt{\frac{2}{3}} \simeq 0.816 \quad\text{for} \quad p=\infty\,.\label{eq:boundD5pinfty}
\end{align}

\subsubsection{$D=7$}

Similarly, the mass-like term of the effective potential in $D=7$ dimensional spacetime reads

\begin{align}
 U_{\kir}^{\text{D=7}}\Big|_{\phi_0^2} =&\frac{\Nf\phi_0^2}{2}\bigg[\frac{1}{\bar{\lambda}_{\Lambda}} -\frac{1}{\blamc}
 +\frac{\Gamma\big(1-\frac{5}{2p}\big)}{320 \pi^{\frac{7}{2}}} \kir^5 \notag\\
    &
    -\frac{15\Gamma\big(1+\frac{1}{2p}\big)}{512\pi^{\frac{7}{2}}}\frac{\kappa^{6}}{\kir}\bigg]\,.  
\label{eq:UKD7}
\end{align}
This time, a range of admissible regulators includes
$p\in[3,\infty]$. Assuming a subcritical coupling, we can again read
off the curvature bounds which for the extremal regulators are given
by

\begin{align}
 \frac{\kappa}{\kir}&\leq0.928\,, \quad\text{for} \quad p=3\,,\label{eq:boundD7p3}\\
 \frac{\kappa}{\kir}&\leq0.689\,, \quad\text{for} \quad p=\infty\,.\label{eq:boundD7pinfty}
\end{align}
\subsubsection{Dimensional dependence}

\begin{figure}
 \includegraphics[width=0.45\textwidth]{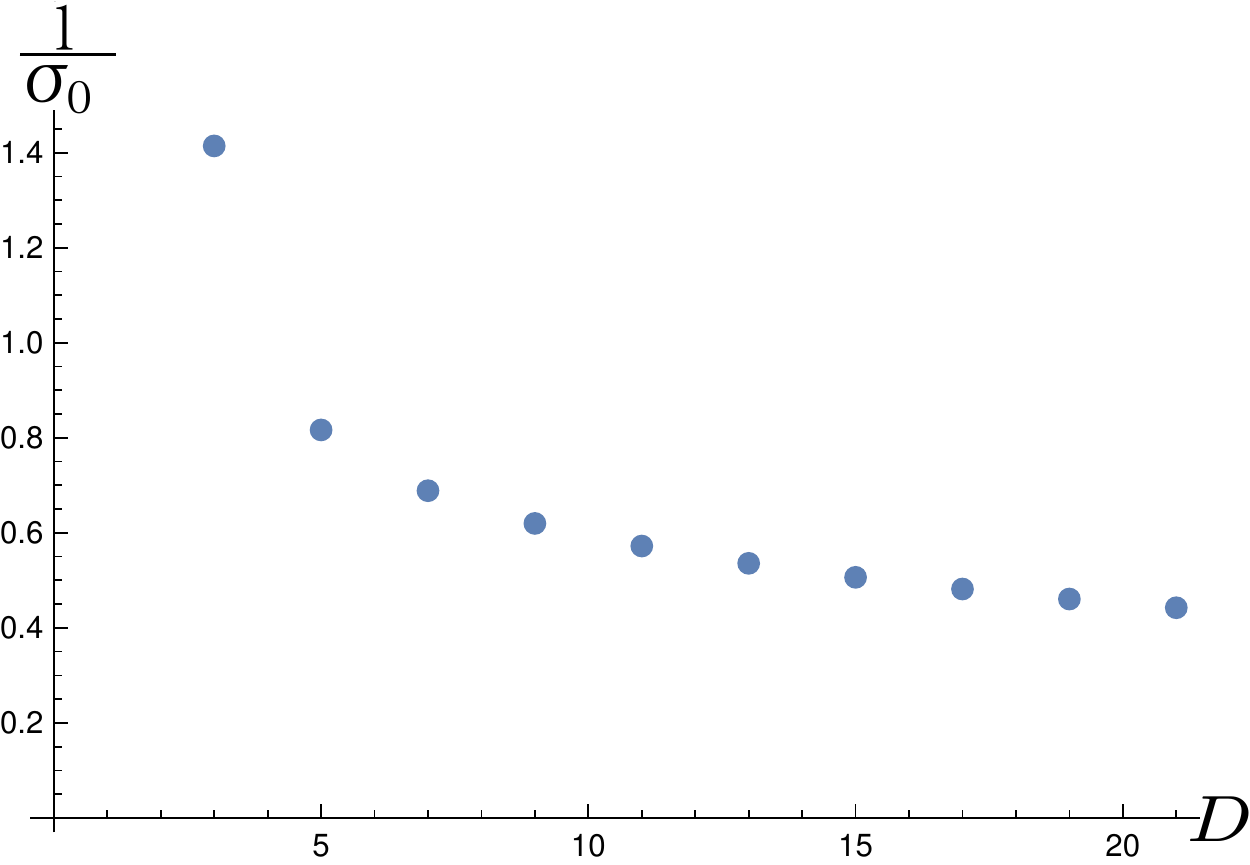}
 \caption{Curvature bound \eqref{eq:boundDgeneral} as a function of
   the spacetime dimensions in the odd-dimensional case for the
   regulator with $p\to\infty$.}
 \label{fig:boundVSdim}
\end{figure}

As is obvious from all these examples, the curvature bound arises from
a competition between the screening of the long-range modes
parametrized by the last term in \Eqref{eq:flatD} and the dominant
curvature term given by \Eqref{regularOpOddDim}. For general $D$, we
need to use the regulator with $p\to\infty$ to ensure that we stay
away from regularization artifacts in any $D$. In order to perform a
meaningful comparison, we set all possible nonzero scalar-curvature
interactions terms $\sim \xi_{\kir,m}$ to zero. For this, the
curvature bound can be expressed as follows:
\begin{align}
 \frac{\kappa}{\kir}\leq\frac{1}{\sigma_0}\equiv\Bigg(\frac{\sqrt{\pi}}{\Gamma\Big(\frac{D}{2}\Big)(D-2)}\Bigg)^{\frac{1}{D-1}}\,,
\label{eq:boundDgeneral}
\end{align}
exhibiting a  monotonically decreasing behavior as is visible in Fig.~\ref{fig:boundVSdim}. Asymptotically, the bound decays as $\sim 1/\sqrt{D}$.

Of course, in the presence of further nonzero scalar-curvature
interaction terms $\sim \xi_{\kir,m}$, the bound can be shifted in
both directions depending on the precise parameter values.

\section{Application: asymptotically safe gravity}
\label{sec:applic}

As an illustration for the application of our curvature bound, we use
a specific quantum gravity scenario in $D=4$ dimensional spacetime:
asymptotically safe gravity
\cite{Weinberg:1976xy,Weinberg:1980gg,Reuter:1996cp,Reuter:2001ag,Litim:2003vp,Niedermaier:2006wt,Reuter:2012id,Percacci:2017book}
In this scenario, Einstein's gravity arises as the low-energy limit of
a quantum field theory of the metric, the high-energy behavior of which
is controlled by a non-Gau\ss{}ian fixed point in the space of
relevant couplings. For simplicity, we confine ourselves to the theory
space spanned by the Einstein-Hilbert action. A more comprehensive
analysis suggests the existence of one further relevant operator with
an overlap with an $R^2$-term in the action \cite{Lauscher:2002sq,Codello:2006in,Machado:2007ea,Benedetti:2009rx,Dietz:2012ic,Falls:2013bv,Falls:2014tra,Demmel:2015oqa,Ohta:2015efa,Falls:2016msz,Christiansen:2016sjn,Denz:2016qks,Gonzalez-Martin:2017gza,Falls:2017lst}. For a first
glance at the consequences of the curvature bound, we also ignore the
influence of the scalar-curvature operator $\sim \phi^2 R$, which is,
in principle, calculable within asymptotic safety from the fermionic
operator content, i.e., schematically $\sim (\bar\psi \psi)^2 R$.

In the modern functional renormalization group approach \cite{Wetterich:1992yh} to
asymptotically safe gravity \cite{Reuter:1996cp}, one studies a scale-dependent
effective action $\Gamma_k$ governing the dynamics of the expectation
values of the field degrees of freedom, typically using a
background-field gauge with a fiducial but arbitrary background
metric. In the simple Einstein-Hilbert truncation, the background
metric itself is a solution to the equation of motion derived from the
scale-dependent effective action \cite{Lauscher:2005qz},
\begin{equation}
  R_{\mu\nu}(\langle g\rangle_k)= \bar{\Lambda}_k \langle g_{\mu\nu}\rangle_k,
  \label{eq:EEscaledep}
\end{equation}
where $\bar{\Lambda}_k$ denotes the scale-dependent cosmological
constant, and $k$ is the coarse-graining or resolution scale, at which
the spacetime is considered. Here, we have assumed the absence of any
explicit matter sources. The asymptotic safety scenario provides us
with a prediction for the RG trajectories for the cosmological
constant $\bar{\Lambda}_k$, as well as for the UV fixed-point value
$\lim_{k\to\infty} \bar{\Lambda}_k/k^2 = \lambda_\ast$ being a finite
number. In the fixed-point regime, the solution to \Eqref{eq:EEscaledep} is given by
\begin{equation}
  \frac{R}{k^2}=4 \lambda_\ast.
  \label{eq:FPregime}
\end{equation}
This shows that the sign of the curvature in the fixed-point regime is
dictated by the sign of the fixed-point value of the cosmological
constant. Equation \eqref{eq:FPregime} exemplifies the self-similarity
property of physical observables in the fixed-point regime: the
curvature is proportional to the scale $k$ at which the curvature is
measured. While the fixed-point value $\lambda_\ast$ comes out
positive in pure-gravity computations, it can change sign for
an increasing number of fermionic degrees of freedom. Hence, the
spacetime structure appears locally as negatively curved for large
$\Nf$. The asymptotic safety scenario including matter degrees of
freedom hence predicts that a local patch of spacetime in the
(trans-Planckian) fixed-point regime satisfies
\begin{equation}
  \frac{\kappa^2}{k^2} = \frac{|\lambda_\ast|}{3}, \quad \text{for}\,\,\,\lambda_\ast<0.
  \label{eq:FPpredict}
\end{equation}
Now, the precise value of the fixed point $\lambda_\ast$ is scheme
dependent, see, e.g.,
\cite{Lauscher:2001ya,Codello:2008vh,Gies:2015tca,Ohta:2016npm,Falls:2017cze,Alkofer:2018fxj} for
comparative studies. With regard to \Eqref{eq:FPregime} this is
natural, since the result of a curvature measurement is expected to
depend on the coarse-graining procedure that is used to average over
metric fluctuations. This is precisely the type of scheme dependence,
we expect to cancel the scheme dependence of our curvature
bounds in order to arrive at a scheme-independent answer to the
question as to whether or not there is gravitational catalysis in a
given theory.

For the remainder of the section, we simply identify the gravitational
RG coarse-graining scale $k$ with the scale $\kir$ used for our
curvature bounds and use results obtained in the asymptotic-safety
literature. In fact, the typical fixed-point scenario can already be
discovered within a simple one-loop calculation
\cite{Codello:2008vh,Percacci:2017book}, yielding the fixed-point values for
the cosmological constant and the dimensionless Newton constant
\begin{equation}
\lambda_\ast = \frac{3}{4} \frac{2+d_\lambda}{46-d_g}, \quad g_\ast = \frac{12\pi}{46 -d_g}.
\end{equation}
Here, we used the results obtained from a so-called type IIa cutoff
\cite{Percacci:2017book}. The two parameters $d_g$ and $d_\lambda$ are
determined by the number of (free) matter degrees of freedom,
\begin{equation}
  d_g=N_{\text{S}}-4N_{\text{V}}+2\Nf, \quad d_\lambda= N_{\text{S}}+2 N_{\text{V}}-4\Nf,
      \label{eq:dgdl}
\end{equation}
where $N_{\text{S}}$ denotes the number of real scalar fields,
$N_{\text{V}}$ the number of gauge vector bosons and -- as before --
$\Nf$ the number of Dirac fermion flavors.

For gravitational catalysis to be potentially active at all, we need a negative fixed-point value $\lambda_\ast<0$, implying
\begin{equation}
\Nf > \frac{1}{2} + \frac{N_{\text{S}}}{4} + \frac{N_{\text{V}}}{2}.
\label{eq:mincriterion}
\end{equation}
This criterion is satisfied for the standard model with
$N_{\text{S}}=4$, $N_{\text{V}}=12$ and $\Nf=45/2$, as well as typical
generalizations with right-handed neutrino components, axion or simple
scalar dark matter models. It is also generically satisfied for
supersymmetric models; for instance, for the MSSM with two Higgs
doublets, we have $N_{\text{S}}=53$, $N_{\text{V}}=12$ and
$\Nf=65/2$. This exemplifies that the curvature bound should be
monitored in asymptotically safe gravity-matter systems. However, the
criterion \eqref{eq:mincriterion} is typically not satisfied for
GUT-like nonsupersymmetric theories where the contribution from larger
number of gauge bosons and Higgs fields for the necessary symmetry
breaking exceeds that of the fermion flavors.

For a given number of scalars and vectors, increasing the number of
flavors drives the fixed point $\lambda_\ast$ towards more negative
values. Using \Eqref{eq:FPpredict} with $k=\kir$, the averaged
curvature can eventually violate the curvature bound. Hence, the
curvature bound translates into an upper bound $\Nf\leq N_{\text{f,gc}}$
on the number of fermion flavors in order not to be inflicted by
chiral symmetry breaking from gravitational catalysis. For instance,
for a purely fermionic matter content, $N_{\text{S}}=0=N_{\text{V}}$,
we find $N_{\text{f,gc}}\simeq 17.58$ for $p\to\infty$, and
$N_{\text{f,gc}}\simeq 18.31$ for $p=2$,
cf. \Eqref{eq:approxboundD4num}. The scheme dependence of our
curvature bound thus has only a mild influence on the critical fermion
number. 

Similarly, fixing the bosonic matter content to that of the standard
model, $N_{\text{S}}=4$, $N_{\text{V}}=12$, the corresponding critical
fermion number is $N_{\text{f,gc}}\simeq 35.97$ for $p\to\infty$. This
would still allow for a fourth generation of standard model flavors,
but exclude a fifth generation. 

Interestingly, the MSSM with $N_{\text{S}}=53$ and $N_{\text{V}}=12$
would imply a critical flavor number of $N_{\text{f,gc}}\simeq 20.3$
far below the fermionic content of the model $\Nf=65/2$, thus 
indicating a possible tension between asymptotically safe gravity and
a particle-physics matter content of that of the MSSM because of
gravitational catalysis.

This analysis based on a simple one-loop calculation on the gravity
side may be somewhat over-simplistic. In fact, a number of more
sophisticated analyses have been performed for asymptotically safe
gravity in conjunction with matter systems. A first study on the
consistency of asymptotic safety with matter \cite{Dona:2013qba} was
based on the background-field approximation with some improvements for
the anomalous dimensions. Using their fixed-point results, we find
$N_{\text{f,gc}}\simeq 8.21$ for a purely fermionic model
($N_{\text{S}}=0=N_{\text{V}}$), and $N_{\text{f,gc}}\simeq 26.5$ for
the standard model model with $N_{\text{S}}=4$, $N_{\text{V}}=12$ (and
anomalous dimensions set to zero). The latter result includes the
standard-model fermion content without and with right-handed neutrino
partners, but does not offer room for a fourth generation. For the
MSSM and other models there is not even a gravitational fixed point
according to \cite{Dona:2013qba}. Even if we artificially reduce the
number of fermion flavors, we do not find a suitable fixed point above
$\Nf\simeq 17$. Here, $\lambda_\ast$ has become negative but the
curvature bound is still satisfied.

The fixed-point scenario found in
\cite{Meibohm:2015twa,Christiansen:2017cxa} is different. The
calculation distinguishes between the background field and the
dynamical fluctuation field. The flow of the dynamical couplings which
is driven by the dynamical correlators \cite{Christiansen:2015rva} is
found to have a gravitational UV fixed point for any matter content
that has been accessible in this study. This scenario hence does not
rule out any particle-physics content from the side of UV
compatibility with quantum gravity. Still, the predictions for the
background-field couplings are qualitatively similar to those of
\cite{Dona:2013qba}. The fixed-point results of \cite{Meibohm:2015twa}
upon insertion into \Eqref{eq:FPpredict} and a comparison with the
curvature bound suggest $N_{\text{f,gc}}\simeq 48.7$ for $p\to \infty$
for a purely fermionic model with $N_{\text{S}}=0=N_{\text{V}}$; for
$p\to2$, the results of \cite{Meibohm:2015twa} lead to
$N_{\text{f,gc}}\simeq 50.9$.

\begin{table}[t] 
\centering
  \begin{tabular}{|c|c|c|c|}
  \hline%\toprule
  $\qquad$ & \multicolumn{3}{|c|}{$\Nfgc$}  \\ \cline{2-4}
  $\qquad$ & $\,$ PF $\,$ & $\,$ SM+$N_{\text{f}}$ $\,$ & $\,$ MSSM+$N_{\text{f}}$ $\,$ \\ \hline
  $\,$ \begin{minipage}{3cm} one-loop approx.\\ (type IIa) \cite{Codello:2008vh,Percacci:2017book}
  \end{minipage}
  $\,$ & 17.58  & 35.97 & 20.3\\ \hline
   \begin{minipage}{3cm} background-field \\ approximation \cite{Dona:2013qba}
   \end{minipage}
   & 8.21 & 26.5 & no FP\\ \hline
   \begin{minipage}{3cm} RG flow on foliated\\ spacetimes \cite{Biemans:2017zca}
   \end{minipage}
   & 9.27 & 27.67 & 10.01\\ \hline
   \begin{minipage}{3cm}dynamical\\ FRG \cite{Meibohm:2015twa}
   \end{minipage}
   & 48.7 &  &  \\
   \hline
  \end{tabular}
  \caption{Summary of the critical number of fermion species $\Nfgc$ below which particle theories are safe from chiral symmetry breaking through gravitational catalysis, using the $p\to\infty$ regulator. Results are shown for theories with a purely fermionic matter content (PF), the standard model and the MSSM artificially varying the number of fermions (SM+$\Nf$, MSSM+$\Nf$). For an estimate of the UV properties of quantum spacetime, we use various literature results obtained within the asymptotic safety scenario of quantum gravity, see main text for details.}
  \label{tab:boundComparison}
\end{table}

Recently, an analysis of gravity-matter systems was performed in
\cite{Biemans:2017zca} using an ADM decomposition of the gravitational
degrees of freedom, yielding an RG flow on foliated spacetimes. For
both, gravitational as well as matter degrees of freedom, a type I
regulator was used. As argued by the authors, the use of different
regulators can be viewed as yielding a different map of the number of
degrees of freedom $N_{\text{S}}$, $N_{\text{V}}$ and $\Nf$ onto the
parameters $d_g$ and $d_\lambda$; e.g., for the type I regulator, one
gets \cite{Biemans:2017zca,Percacci:2017book}
$d_g=N_{\text{S}}-N_{\text{V}}-\Nf$. It has been argued that
the type II regulator should be used for fermions in order to regulate
the fluctuation spectrum of the Dirac operator in a proper fashion
\cite{Dona:2012am,Percacci:2017book}. Hence, we use the flows of
\cite{Biemans:2017zca} but with a definition of the parameters $d_g$
and $d_\lambda$ as in \Eqref{eq:dgdl}. To leading order, this corresponds to a
type I regularization of the gravity fluctuations but a type II
regulator for the matter degrees of freedom.

In this case, the possible onset of gravitational catalysis for a
purely fermionic model with $N_S=0=N_V$ occurs at a critical flavor
number $\Nfgc=9.27$ for $p\to\infty$ ($\Nfgc=9.84$ for $p=2$).  For a
standard-model like theory ($N_S=4$, $N_V=12$), we have $\Nfgc=27.67$
$p\to\infty$ ($\Nfgc=28.71$ for $p=2$). Finally, the minimally
supersymmetric extension of the standard model would lead to
$\Nfgc=10.01$ for $p\to\infty$ ($\Nfgc=10.27$ for $p=2$), if we
artificially allow $\Nf$ to vary independently in this
model. Therefore the MSSM in this approximation is an example for a
model where gravitational catalysis could lead to large-fermion-mass
generation in the trans-Planckian regime; in fact, if $\Nf$ is set to
the physical value $\Nf=65/2$, the MSSM matter content in this setting
does not lead to a fixed point suitable for asymptotically safe
quantum gravity, see also \cite{Alkofer:2018fxj}.

We summarize the critical values for the fermion numbers $\Nfgc$ for
$p\to\infty$ for the possible onset of gravitational catalysis derived
within the various approximations for an asymptotically safe quantum
gravity scenario in Tab.~\ref{tab:boundComparison}. Whereas the
standard model (e.g., also including right-handed neutrinos) satisfies
the bound from gravitational catalysis in each of thee approximations,
a standard model with a fourth fermion generation could already be
affected by gravitational catalysis. Supersymmetric versions of the
standard model show already some tension with the bound within
asymptotically safe gravity.

\begin{figure}
 \includegraphics[width=0.5\textwidth]{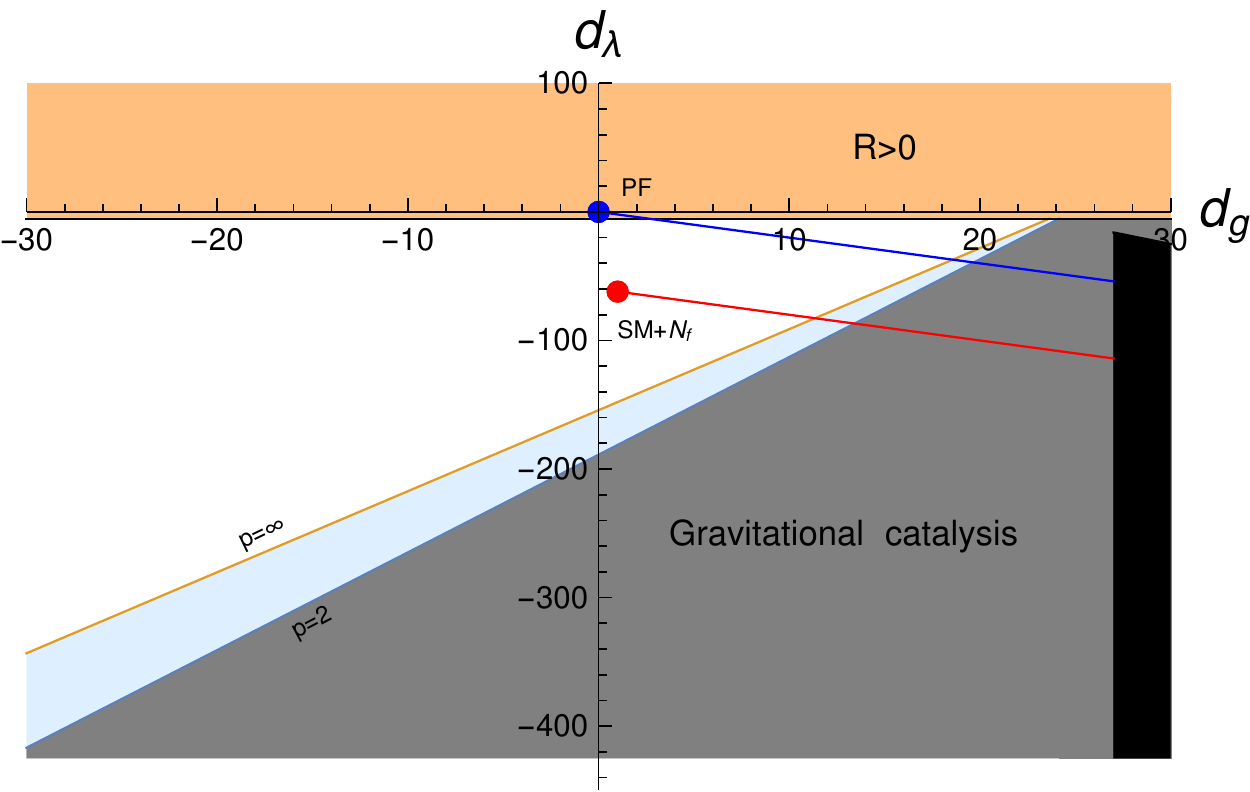}
 \caption{Relevance of gravitational catalysis in different regions in
   the space of matter theories parametrized by $d_g$ and $d_\lambda$,
   cf. \Eqref{eq:dgdl}, using the results of \cite{Biemans:2017zca} as
   an input from the asymptotic safety scenario for quantum
   gravity. In the bright orange region in the upper part, the criterion
   analogous to \eqref{eq:mincriterion} is not satisfied, indicating
   that gravitational catalysis does not occur imposing no
   constraints. Our curvature bound results in a line with the (white)
   region above that line satisfying the bound; we observe a mild
   regulator dependence for different regulator parameters
   $p\in[2,\infty]$.  Further lines show the location of purely
   fermionic particle models (PF) for an increasing number of flavors
   (blue), as well as the standard model (red dot) with additional
   flavor numbers (SM+$\Nf$). The gray region at the bottom indicates
   the region where chiral symmetry breaking and fermion mass
   generation because of gravitational catalysis can occur. The black
   region on the right does not have a non-Gau\ss{}ian fixed point
   suitable for asymptotic safety \cite{Biemans:2017zca}.}
 \label{fig:dGdLambdaPlane}
\end{figure}

Using the results of \cite{Biemans:2017zca} as described above, we
display the various regions in the space of matter theories
parametrized by $d_g$ and $d_\lambda$, cf. \Eqref{eq:dgdl}, in
Fig.~\ref{fig:dGdLambdaPlane}. In the upper orange-shaded region, the
criterion analogous to \eqref{eq:mincriterion} is not satisfied (in
the calculation of \cite{Biemans:2017zca}, it corresponds to
$d_\lambda> -16/3$); here, we expect a spacetime in the fixed-point
regime which is positively curved and thus not affected by
gravitational catalysis. The curvature bound translates into a line in
the $d_g,d_\lambda$ plane, with the (white) region above that line
satisfying the bound. We observe that the lines for different
regulators $p\in[2,\infty]$ are rather similar and deviate
significantly only for extreme particle numbers. The purely fermionic
model (PF) and the standard model are represented by dots in the
plane. The lines attached to the dots correspond to increasing the
fermion number in these models. The purely fermionic model starts at
$\Nf=0$, while the standard-model starts at its physical value
$\Nf=45/2$. The MSSM with $\Nf=65/2$ would lie deep inside the black
region to the right where no fixed point suitable for asymptotically
safe gravity exists \cite{Biemans:2017zca,Alkofer:2018fxj}.

Let us close this section with two remarks: the first remark concerns
the regularization scheme dependence which occurs at various places in
this calculation. In case of a fully consistent calculation this
scheme dependence would cancel in the final result for
$\Nfgc$. However, since different parts in the present estimates are
performed with different regulators, we observe various sources of
scheme dependence. Whereas the scheme dependence arising from our
mean-field RG calculation parametrized by $p$ is rather mild, a change
of the regulator from type II to type I in the asymptotic safety
scenario can change the dependence on the fermion flavor content
significantly as studied in the literature
\cite{Dona:2012am,Percacci:2017book}. Since our fermionic mean-field
RG calculation corresponds to a type II regularization, we find it
reassuring that a consistent use of type II regulators for the
fermions leads to qualitatively and partly quantitatively similar
results in the various approximations.

Second, the asymptotic safety scenario suggests that at least one
further relevant operator of $R^2$ type should be included in the
fixed-point regime. As this could take a significant quantitative
influence on the effective equation of motion in the fixed-point
regime, cf. \Eqref{eq:EEscaledep}, the relevance of the curvature
bound for the asymptotic safety scenario may also change
qualitatively. With these reservations in mind, the present discussion
should be viewed as an example how the curvature bound from
gravitational catalysis could potentially be used to constrain
combined scenarios of quantum gravity and quantum matter.

\section{Conclusions}
\label{sec:conc}

We have studied gravitational catalysis of chiral symmetry breaking
and fermion mass generation on patches of hyperbolic spaces,
corresponding to negatively curved patches of AdS spacetimes in a
Lorentzian setting. The general phenomenon of gravitationally
catalyzed symmetry breaking has long been known to be driven by
long-range modes and their sensitivity to the large-scale structure of
negatively curved spacetimes. In this work, we have analyzed for the
first time the competition between the screening of these modes by a
gauge-invariant IR averaging scale $\kir$ and the effect of the
presence of an averaged curvature on this scale. This competition
leads to a bound on the local curvature parameter $\kappa \sim
\sqrt{|R|}$ in units of the averaging scale $\kir$. Gravitational
catalysis does not set in as long as the bound is satisfied.

Built on RG-type arguments, our analysis applies to local patches of
spacetime and hence does not require the whole spacetime to be
hyperbolic, negatively curved or uniform. Rather the resulting bound
applies to each patch of space or spacetime with an averaged negative
curvature. Fermion modes in spacetime patches violating the bound can
be subject to gravitational catalysis. The precise location of the
onset of gravitational catalysis in parameter space depends also on
further induced or fundamental interactions of the fermions. In case
of chiral symmetry breaking through gravitational catalysis, the
fermions generically acquire masses of the order of at least $\kir$ or
larger depending on the relevance of further effective interactions.

An application of these findings to a possible high-energy regime of
quantum gravity results in the following scenario: let us assume the
existence of a, say Planck scale, regime where a metric/field theory
description is already appropriate, but large curvature fluctuations
are allowed to occur. Our bound disfavors the occurrence of patches of
spacetime with large negative averaged curvature. In such patches, the
generation of fermion masses of the order of $\kir$ could be
triggered. Since $\kir$ itself can be of order Planck scale in such a
regime, the fermion masses would generically be at the Planck scale
upon onset of gravitational catalysis. Even worse, gravitational
catalysis would naturally remove light fermions from the spectrum of
particle physics models on such spacetimes. Therefore, we argue that
our bounds apply to any quantum gravity scenario satisfying these
assumptions that aims to be compatible with particle physics
observations: if a quantum gravity scenario satisfies the bound, it is
safe from gravitational catalysis in the matter sector; if not, the
details of the fermion interactions matter. In the latter case,
gravitational catalysis may still be avoided, if the interactions
remain sufficiently weak.

As the curvature bounds refer to an IR cutoff scale $\kir$, they are
naturally scheme dependent. In fact, this scheme dependence in the
first place parametrizes the details of how the fermionic long-range
modes are screened by the regularization scale. We observe that a
finite curvature bound exists for any physically admissible
regularization. Moreover, the shifts of the bound due to a change of
the propertime regularization agrees with the behavior expected from
the underlying propertime diffusion process. We therefore claim that
the curvature bound has a scheme-independent meaning. A fully
scheme-independent definition might eventually need to take the
prescription for defining the averaged curvature of a local spacetime
patch into account.

Having performed a mean-field-type RG analysis, our bounds may receive
corrections from further fluctuations that may be relevant at the
scale $\kir$ including further independent degrees of freedom or
chiral-order-parameter fluctuations. Such corrections can go into both
directions: further interactions such as gauge or Yukawa forces
typically enhance the approach to chiral symmetry breaking, whereas
order-parameter fluctuations can have the opposite effect. Also,
thermal fluctuations can inhibit the occurrence of a chiral condensate
at sufficiently high temperature. Effects that trigger
symmetry-breaking can effectively be summarized in terms of finite
bare fermionic self-interactions $\blamL$ in our approach, whereas
thermal fluctuations can be understood as moving the critical coupling
to larger values \cite{Braun:2006jd,Braun:2011pp}.

We have been able to determine the curvature bound also in $D=3$ as
well as in higher dimensions. In general odd dimensions, we have
derived a simple closed form expression. Since different dimensions
can exhibit a different number of relevant scalar-curvature operators
and thus a different number of physical parameters, a meaningful
comparison of theories in different dimensions is not
straightforward. Assuming that all further physical parameters are
essentially zero at the scale $\kir$, we observe that the resulting
curvature bound decreases with $\sim 1/\sqrt{D}$ for higher
dimensions. This result inspires to develop the following scenario:
Let us assume that some fundamental theory of spacetime and matter can have
a high-energy phase of arbitrary dimension and allows for a regime
where a metric description applies. If the theory in addition exhibits
fluctuating values of curvature $\kappa \sim \mathcal{O}(1)$ when
averaged over local patches, our results suggest that it is unlikely
to find higher-dimensional regions that admit massless or light
fermions in the long-range physics. Upon the onset of gravitational
catalysis, higher-dimensional regions would then generically go along
with a massive fermionic particle content and without explicit chiral
symmetry.

Unfortunately, results from quantum gravity scenarios that could be
checked against our curvature bounds are rather sparse. Many
approaches focus on the gravitational sector leaving matter, and
fermions in particular, aside. One of the most developed approaches in
this respect is asymptotically safe gravity. Concentrating on a simple
picture for the UV regime of gravity using the Einstein-Hilbert action
as the scaling action, our curvature bound translates into a bound on
the particle content of the matter sector. In particular, the number
of fermion flavors becomes constrained in order to avoid gravitational
catalysis. Our simple estimates based on various literature studies of
asymptotically safe gravity with matter indicate that the standard
model is compatible with asymptotically safe gravity and not affected
by gravitational catalysis in the trans-Planckian regime. This
statement is nontrivial insofar that the matter content together with
the effective Einstein equation suggest negatively curved local
patches of spacetime in the fixed-point regime. Still, the curvature
is sufficiently weak to satisfy our curvature bound. By contrast, our
estimates suggest that the standard model with an additional fourth
flavor generation would not satisfy the curvature bound within
asymptotic safety. In order to obtain more reliable estimates, the
curvature dependence of correlation functions and its interdependence
with the matter sector in the trans-Planckian fixed-point regime would
be welcome.

This first application within a specific quantum gravity scenario
demonstrates that our curvature bound may be usefully applicable also
in the high-energy regime of other quantum gravity scenarios.

%%%%%%%%%%%%%%%%%%%%%%%%%%%%%%%%%%%%%%%%
\section*{Acknowledgments}
%%%%%%%%%%%%%%%%%%%%%%%%%%%%%%%%%%%%%%%%

We thank Astrid Eichhorn, Benjamin Knorr, Stefan Lippoldt, Jan
Pawlowski, Frank Saueressig, Manuel \mbox{Reichert}, and Omar Zanusso for
valuable discussions. We are grateful to Astrid Eichhorn, Frank
Saueressig, and Manuel Reichert for making their data and Mathematica
files available to us. We acknowledge support by the DFG under Grants
No. GRK1523/2 and No. Gi328/7-1.

\appendix

\section{Heat kernel on hyperbolic spaces}
\label{app:heatKer}

For completeness, we summarize results for the heat kernel on
hyperbolic spaces in this appendix, as they are needed for the present
work. Following the derivation and conventions of
\cite{Camporesi:1992tm,Camporesi:1995fb}, we normalize the inverse
radius $\kappa$ of the manifold to 1, and reinstate this curvature
parameter later. The heat kernel on a $D$-dimensional hyperbolic space
can be written as:
\begin{align}
\label{generalHK}
 K(x, x', T) &= \mathbb{U}(x, x')\hat{f}_N(d_G, T)\,,\\
 \label{generalHKScalar}
 \hat{f}_N(y, T) &= \frac{2^{D-3}\Gamma\Big(\frac{D}{2}\Big)}{\pi^{\frac{D}{2}+1}}
      \int_0^{\infty}\varphi_{\lambda}(y)e^{-T\lambda^2}\mu(\lambda)d\lambda\,,
\end{align}
where $\mathbb{U}$ represents a parallel transport, and $\hat{f}_N$ is a scalar
function satisfying the following equation:
\begin{align}
 0 &= \Big(-\frac{\partial}{\partial T}+\Box_D-\frac{R}{4}-\frac{D-1}{4}\tanh^2(y)\Big)\hat{f}_N(y)\\
 &\equiv \Big(-\frac{\partial}{\partial T}+L_D\Big)\hat{f}_N(y)\,,
 \label{eq:fhat}
\end{align}
with $\Box_D$ being the radial Laplacian. The eigenfunctions $\varphi_{\lambda}$  of the $L_D$ operator  with eigenvalues $-\lambda^2$ can be written as
\begin{align}
 L_{D}\varphi_{\lambda} &= -\lambda^2\varphi_{\lambda}
 \label{LnEigenfunctions}\\
 \varphi_{\lambda}(y) &= \cosh\frac{y}{2}\, {}_{2}F_1\Big(\frac{D}{2}+i\lambda, \frac{D}{2}-i\lambda; \frac{D}{2};-\sinh^2 \frac{y}{2}\Big). \notag
\end{align}
Here, ${}_{2}F_1$ denotes the hypergeometric function, while the
spectral measure $\mu(\lambda)$ reads:
\begin{align}
 \mu(\lambda) =&\frac{\pi}{2^{2D-4}\Gamma^2\Big(\frac{D}{2}\Big)} \label{eq:specmeasure}\\
 &  \times   \begin{cases}
	    \prod_{j=\frac{1}{2}}^{\frac{D}{2}-1}(\lambda^2+j^2)\,,\quad D\quad\text{odd}\\
	    \lambda\coth(\pi\lambda)\prod_{j=1}^{\frac{D}{2}-1}(\lambda^2+j^2)\,,\; D\;\text{even}\,.
       \end{cases}
 \notag
\end{align}
In the main text, cf. Sect.~\ref{sec:framework}, we only need the
equal point limit of the heat kernel, with $x'\to x$ and the geodesic
distance $d_G\to 0$ goes to zero, i.e., $y \to 0$ in
\eqref{generalHKScalar}. From equation \eqref{LnEigenfunctions} is
clear that the coincident points limit leads to
\begin{align}
 \lim_{y\to 0} \varphi_{\lambda}(y) = 1\,, \label{eq:coincl}
\end{align}
while the $\mathbb{U}$ reduces to the identity. Thus, we end up with
\begin{align}
\label{generalHKcoincidentLimit}
 K_T = \frac{2^{D-3}\Gamma\Big(\frac{D}{2}\Big)}{\pi^{\frac{D}{2}+1}}
      \int_0^{\infty}d\lambda\; e^{-T\lambda^2}\mu(\lambda)d\lambda\,.
\end{align}

In order to reinstate the curvature parameter, we make contact with the flat space limit of the heat kernel, starting with the odd dimensional
case. Plugging the definition of $\mu(\lambda)$ into equation
\eqref{generalHKcoincidentLimit}, we get upon substitution
\begin{align}
   K_T^{\text{odd}} &= \frac{2}{(4\pi)^{\frac{D}{2}}\Gamma\Big(\frac{D}{2}\Big)}\int_0^{\infty}d\lambda\;e^{-T\lambda^2}
	   \prod_{j=\frac{1}{2}}^{\frac{D}{2}-1}(\lambda^2+j^2)\notag\\
%   &= \frac{2}{(4\pi T)^{\frac{D}{2}}\Gamma\Big(\frac{D}{2}\Big)}\int_0^{\infty}d\lambda\sqrt{T}\;e^{-T\lambda^2}
%	   \prod_{j=\frac{1}{2}}^{\frac{D}{2}-1}(\lambda^2T+j^2T)\\
   &= \frac{2}{(4\pi T)^{\frac{D}{2}}\Gamma\Big(\frac{D}{2}\Big)}\int_0^{\infty}du\;e^{-u^2}
	   \prod_{j=\frac{1}{2}}^{\frac{D}{2}-1}(u^2+j^2T)\,,
 \label{HKodd}
\end{align}
and similarly for an even dimensional background:
\begin{align}
  K_T^{\text{even}}
  %=& \frac{2}{(4\pi)^{\frac{D}{2}}\Gamma\Big(\frac{D}{2}\Big)}\int_0^{\infty}d\lambda\;e^{-T\lambda^2}
%	   \lambda\coth(\pi\lambda)\prod_{j=1}^{\frac{D}{2}-1}(\lambda^2+j^2)\\
%=& \frac{2}{(4\pi T)^{\frac{D}{2}}\Gamma\Big(\frac{D}{2}\Big)}\int_0^{\infty}d\lambda\sqrt{T}\;e^{-T\lambda^2}\lambda\sqrt{T}\coth(\pi\lambda)\\
%&\prod_{j=1}^{\frac{D}{2}-1}(\lambda^2T+j^2T)\\
=& \frac{2}{(4\pi T)^{\frac{D}{2}}\Gamma\Big(\frac{D}{2}\Big)}\int_0^{\infty}du\;e^{-u^2}
u\coth(\pi\frac{u}{\sqrt{T}})\notag\\
&\qquad\times\prod_{j=1}^{\frac{D}{2}-1}(u^2+j^2T)\,.
 \label{HKeven}
\end{align}
Recalling that in flat spacetime the heat kernel in the coincident
points limit reads $K_T = (4\pi T)^{-1}$ with $T$ carrying mass
dimension $[T] = -2$, we obtain the correct limit by rescaling the
propertime inside the integrals by a sufficient power of the curvature
parameter with $[\kappa] = 1$; note that the integration variables has
to remain dimensionless, $[u]=0$. We finally obtain,
\begin{align}
\label{generalHKodd}
  K_T^{\text{odd}} =& \frac{2}{(4\pi T)^{\frac{D}{2}}\Gamma\Big(\frac{D}{2}\Big)}\int_0^{\infty}du\;e^{-u^2}
	   \prod_{j=\frac{1}{2}}^{\frac{D}{2}-1}(u^2+j^2\kappa^2T)\,,\\
  \label{generalHKeven}
  \begin{split}
  K_T^{\text{even}} =& \frac{2}{(4\pi T)^{\frac{D}{2}}\Gamma\Big(\frac{D}{2}\Big)}\int_0^{\infty}du\;e^{-u^2}
	  u\coth(\pi\frac{u}{\kappa\sqrt{T}})\\
  &\times\prod_{j=1}^{\frac{D}{2}-1}(u^2+j^2\kappa^2T)\,.
  \end{split}
\end{align}
For an analytical approximation  in even dimensions, the
expansion of the integrand in the two limits
$T\approx 0$ and $T\approx\infty$ are useful. For small $T$, we rewrite the
hyperbolic cotangent as:
\begin{align}
\label{cothSmallT}
\coth(\pi\frac{u}{\kappa\sqrt{T}})
%=& \frac{e^{\frac{\pi u}{\kappa\sqrt{T}}}+e^{-\frac{\pi u}{\kappa\sqrt{T}}}}{e^{\frac{\pi u}{\kappa\sqrt{T}}}-e^{-\frac{\pi u}{\kappa\sqrt{T}}}} =
%  \frac{1+e^{-2\frac{\pi u}{\kappa\sqrt{T}}}}{1-e^{-2\frac{\pi u}{\kappa\sqrt{T}}}}\\
% =&\Big(1+e^{-2\frac{\pi u}{\kappa\sqrt{T}}}\Big)\sum_{n=0}^{\infty}e^{-2\frac{\pi n u}{\kappa\sqrt{T}}}\\
%=&\sum_{n=0}^{\infty}e^{-2\frac{\pi n u}{\kappa\sqrt{T}}} + \sum_{n=0}^{\infty}e^{-2\frac{\pi (n+1) u}{\kappa\sqrt{T}}}\\
=&1+2\sum_{n=1}^{\infty}e^{-2\frac{\pi n u}{\kappa\sqrt{T}}}\,.
\end{align}
The large $T$ regime corresponds to the small $u$ approximation of the
hyperbolic cotangent, thus, it suffices to consider the first few
terms in the Laurent expansion of $\coth(\pi\frac{u}{\kappa\sqrt{T}})$
in order to capture the behavior of $K_T$ for $T$ around infinity,
\begin{align}
\label{cothLargeT}
 \coth(\pi\frac{u}{\kappa\sqrt{T}}) = \frac{\kappa\sqrt{T}}{\pi u} + \frac{\pi u}{3\kappa\sqrt{T}} + \mathcal{O}(u^3)\,.
\end{align}
These two approximations are combined in section \ref{sec:fourDim} to
identify an analytic approximation for the heat-kernel trace in four
dimensions.

\bibliography{bibliography}

\end{document}